\documentclass[showpacs,twocolumn,superscriptaddress]{revtex4}

\usepackage{graphicx}
\usepackage{dcolumn}

\usepackage{amsmath}
\usepackage{amssymb}
\usepackage{bm}
\usepackage{color}
\usepackage[normalem]{ulem}
\usepackage[dvipsnames]{xcolor}
\usepackage{hyperref}
\hypersetup{
	colorlinks=true, 
	linkcolor=blue,  
	citecolor=cyan,  
}
\newcommand{\beqn}{\begin{eqnarray}}
\newcommand{\eeqn}{\end{eqnarray}}
\newcommand{\aap}{Astronomy \& Astrophysics}

\begin{document}
	
\title{Observed jet power and radiative efficiency of black hole candidates in Kerr + PFDM model}

	\author{Bakhtiyor Narzilloev}
\email[]{nbakhtiyor18@fudan.edu.cn}
        \affiliation{Center for Astronomy and Astrophysics, Center for Field Theory and Particle Physics, and Department of Physics,
Fudan University, Shanghai 200438, China}
	\affiliation{Ulugh Beg Astronomical Institute, Astronomy St.  33, Tashkent 100052, Uzbekistan}
	 \affiliation{Institute of Fundamental and Applied Research, National Research University TIIAME, Kori Niyoziy 39, Tashkent 100000, Uzbekistan}	

	%
	\date{\today}
\author{Ahmadjon~Abdujabbarov}
	\email[]{ahmadjon@astrin.uz}
	\affiliation{Ulugh Beg Astronomical Institute, Astronomy St.  33, Tashkent 100052, Uzbekistan}
    \affiliation{University of Tashkent for Applied Sciences, Str. Gavhar 1, Tashkent 100149, Uzbekistan}

\author{Bobomurat Ahmedov}
\email[]{ahmedov@astrin.uz}
\affiliation{Ulugh Beg Astronomical Institute, Astronomy St. 33, Tashkent 100052, Uzbekistan}
\affiliation{Institute of Fundamental and Applied Research, National Research University TIIAME, Kori Niyoziy 39, Tashkent 100000, Uzbekistan}

\author{Cosimo Bambi}
\email[]{bambi@fudan.edu.cn}
\affiliation{Center for Astronomy and Astrophysics, Center for Field Theory and Particle Physics, and Department of Physics,
Fudan University, Shanghai 200438, China}
\affiliation{School of Natural Sciences and Humanities, New Uzbekistan University, Tashkent 100007, Uzbekistan}

\begin{abstract}
\textcolor{black}{In this research, we explore the electromagnetic energy emitted by astrophysical black holes within the Kerr+PFDM spacetime, a model encompassing rotating black holes surrounded by dark matter. Our investigation focuses on black holes within X-ray binary systems, namely GRS~1915+105, GRO~J1655-40, XTE~J1550-564, A0620-00, H1743-322, and GRS~1124-683. Our findings indicate that the Kerr+PFDM spacetime can account for the radiative efficiency of these sources as determined through the continuum fitting method (CFM). Additionally, employing the Blandford-Znajeck mechanism, we demonstrate the ability to replicate the observed jet power. By combining the outcomes of both analyses for the selected objects, we establish more rigorous constraints on the spacetime parameters. Notably, our results reveal that similar to the Kerr spacetime, the Kerr+PFDM spacetime cannot simultaneously account for the observed jet power and radiative efficiency of GRS~1915+105.}
\end{abstract}
\pacs{04.20.-q, 04.70.-s, 04.70.Bw}

\maketitle

\section{Introduction}

Relativistic jets are commonly observed in active galactic nuclei and black hole X-ray binaries (microquasars). In the case of black hole X-ray binaries, we observe two kinds of jets~\cite{ref92}. {\it Steady jets} show up in the hard state, over a wide range of luminosity of the source, and are normally mildly relativistic. {\it Transient jets} appear in low-mass black hole X-ray binaries when the source switches from the hard to the soft state and its luminosity is close to the Eddington limit; they are normally observed as blobs of plasma moving outward at very high velocities and are thought to be generated near the black hole event horizon. As proposed in Ref.~\cite{ref84}, we may interpret the transient jets as powered by the rotational energy of the black hole and, since they occur at a well defined luminosity, they may be used as standard candles.

Our current understanding of the Universe is that about 25\% of its energy is made of non-baryonic dark matter (DM)~\cite{Bergstrom:2000pn,Bertone:2004pz,Feng:2010gw}. Its exact nature is currently unknown, but in most proposals, DM would be made of weakly interacting particles beyond the Standard Model of Particle Physics. In Ref.~\cite{Chan2022gqd}, the authors claim that the (otherwise unexplained) anomalous orbital decay of some low-mass black hole X-ray binaries could be interpreted as the existence of a DM halo around those stellar-mass black holes since the orbital decay could be caused by the dynamical friction of the stellar companion in the DM halo.

In this paper, we study the possible impact of DM around stellar-mass black holes in low-mass X-ray binaries on two observational quantities: the radiative efficiency of the source and the power of relativistic jets. We employ the so-called Kerr+PFDM model (Kerr + perfect fluid dark matter model). As we will show below, the presence of DM surrounding a black hole can alter the position of the innermost stable circular orbit (ISCO) and, in turn, the thermal spectrum of the black hole accretion disk. At the same time, the DM halo will also affect the angular velocity of the event horizon and, in turn, the power of transient jets. The idea to combine the results of these two observations to investigate the spacetime geometry around black hole candidates is not novel and has been explored in a few publications. For example, the relation between jet power and spin was originally found in \cite{ref84} and \cite{ref91}. The extension of this technique to test the spacetime geometry around black holes was discussed in Refs.~\cite{ref78,Banerjee2020ubc,Narzilloev2023h,PhysRevD86123013}. We also note that the properties of various black hole solutions have been extensively studied in our previous works through different methods \cite{Hakimov17,Narzilloev21c,Narzilloev21d,Narzilloev22a,Narzilloev23,Narzilloev22c,Narzilloev23a,Narzilloev2023b,Narzilloev2023c,Narzilloev2023d,Narzilloev2023f,Narzilloev2023g}.

The present paper has the following structure. In Section~\ref{section4.2}, we briefly review the Kerr+PFDM metric and its main properties. In Section~\ref{section4.3}, we give theoretical aspects of observables that can be used to get constraints on the spacetime parameters. In Section~\ref{section4.4}, we get constraints on the spacetime parameters of the Kerr+PFDM metric for selected black hole candidates. We summarize our main results in Section~\ref{section4.6}. Throughout the paper, we use natural units in which $G=c=1$.

\section{Kerr+PFDM spacetime \label{section4.2}}

\textcolor{black}{For a black hole enveloped by dark matter (DM), the DM field interacts with gravity, and the corresponding action is expressed as
\begin{equation}\label{S}
   S=\int d^4x \sqrt{-g} \left(\frac{1}{16 \pi} R+\mathcal{L}_{OM}+\mathcal{L}_{DM}\right),
\end{equation}
where $\mathcal{L}_{OM}$ and $\mathcal{L}_{DM}$ represent the Lagrangian density of ordinary matter and DM, respectively. It's important to note that \eqref{S} does not incorporate the interaction between DM and ordinary matter fields. By employing a variational approach, one can derive the field equation as
\begin{equation}\label{Efe}
   R_{\mu\nu}-\frac{1}{2}g_{\mu\nu}R=-8\pi(T_{\mu\nu}^{OM}+T_{\mu\nu}^{DM}),
\end{equation}
where $T_{\mu\nu}^{OM}$ denotes the energy-momentum tensor of ordinary matter, and $T_{\mu\nu}^{DM}$ is the energy-momentum tensor of dark matter. In the scenario of black holes surrounded by DM, assuming that DM behaves as a perfect fluid, the energy-momentum tensor can be expressed as $T^\mu_\mu=diag [-\rho,p,p,p]$. Additionally, under the simplest conditions, we assume $\gamma - 1 = p/\rho$, where $\gamma$ is a constant \cite{Li2012zx}. This type of DM is referred to as ``perfect fluid DM'' (PFDM).}

\textcolor{black}{In the case of a spherically symmetric spacetime metric involving PFDM, Li et al. presented a solution for the black hole \cite{Li2012zx}. The metric is expressed as:
\[ ds^2 = -f(r) dt^2 + f^{-1}(r) dr^2 + r^2(d\theta^2 + \sin^2\theta d\phi^2), \]
where
\[ f(r) = 1 - \frac{2M}{r} + \frac{\alpha}{r} \ln\left(\frac{r}{|\alpha|}\right), \]
$M$ denotes the black hole mass, and $\alpha$ is some parameter related to the PFDM.} The DM energy density, in this case, takes the following form \cite{Liang2023jrj} $$\rho_{DM}=\frac{3 \alpha}{16 \pi r^3}\, ,$$ so one can interpret $\alpha$ parameter as 'dark matter density' or just intensity parameter of PFDM. From the above relation one can see that the density of DM goes down as $r^3$ with the increase of the distance from the central black hole.

It is worth noting that one might inquire about the possibility of replacing dark matter with a corona having the same energy density. However, a few points need clarification. Firstly, as is known, dark matter does not interact with electromagnetic radiation (besides its gravitational field) whereas corona around black hole is assumed to be consisted from hot electron gas which do interacts with electromagnetic radiation and it is believed that due to inverse Compton scattering can form so called Power Law component of electromagnetic radiation \cite{Bambi17e}. This feature of the corona should make these two types of mediums distinguishable in observations. However, in terms of gravitational effects only, if these two mediums have the same stress-energy tensor distribution, they must be indistinguishable. In this context, it would be more beneficial to focus on studies that examine the gravitational effects of the medium exclusively, where the nature of the medium itself is not a factor. Secondly, we know that the mass of the corona is completely negligible while we do not know the mass of the PFDM, so we can speculate it is high enough to have observable effects. The mass of accretion disks around stellar-mass black holes is at the level of $\sim10^{-8}$ of the mass of the black hole, and the mass of the corona is certainly several orders of magnitudes lower than the mass of the accretion disk. The mass density of dark matter in the Milky Way halo is also very low ($\sim 0.3\, GeV/cm^3$ in the Solar System \cite{Bovy2012tw}) but, we can speculate that the dark matter density is much higher around a compact object like a black hole without disagreements with current observations.

\textcolor{black}{In~\cite{Xu2017bpz} has been obtained the solution for a rotating black hole, referred to as the Kerr+PFDM solution, that describes a black hole with a spin  $\chi$ and intensity of PFDM $\alpha$. The metric is given by
\[ 
\begin{aligned}
ds^2 =& \frac{\Sigma}{\Delta} dr^2 + \Sigma d\theta^2 + \frac{\sin^2\theta}{\Sigma} (\chi dt - (r^2 + \chi^2) d\phi)^2\\ &- \frac{\Delta}{\Sigma} (dt - \chi \sin^2\theta d\phi)^2, 
\end{aligned}
\]
where $\Delta$ and $\Sigma$ are defined as
\[
\begin{aligned}
    \Delta &= r^2 + \chi^2 - 2Mr + \alpha r \ln{\frac{r}{|\alpha|}}, \\
    \Sigma &= r^2 + \chi^2 \cos^2\theta.
\end{aligned}
\]
The horizon radius is determined by the condition $\Delta=0$, expressed as
\begin{equation}\label{ehe}
    r^2 - 2M r + \chi^2 = \alpha r \ln{\frac{r}{|\alpha|}}.
\end{equation}
One can show that the given equation involves two horizons, the inner (Cauchy) horizon $r_-$ and the outer (event) horizon $r_+$~\cite{Xu2017bpz}. 
Throughout this work, we use the dimensionless rotation parameter $a = \chi/M$ and dimensionless intensity parameter $k=\alpha/M$ and consider the latter to be positive $k$ but to be in the order of $\sim0.1$. }

\section{theoretical aspects  \label{section4.3}}

\subsection{Radiative efficiency of the system}

\textcolor{black}{In this section, we explore the continuum spectrum released by an accretion disc around a black hole, focusing on the Novikov-Thorne model~\cite{ref70}. In this model, the accretion disk is taken to be geometrically thin, where particles follow quasi-circular trajectories near the equatorial plane. Viscous forces induce slight radial motions, causing particles to navigate spiral-like paths before eventually succumbing to gravitational attraction and falling into the black hole. }

\textcolor{black}{Given that the gravitational force is presumed to exert a more significant influence on gas motion than gas pressure, we posit that particles within the disc adhere to circular geodesic orbits. As the gas descends into the gravitational well of the central massive object, it undergoes a reduction in both energy and angular momentum. Consequently, a fraction of this lost energy is transformed into electromagnetic radiation.}

\textcolor{black}{The Novikov-Thorne accretion disc is characterized by being geometrically thin, optically thick, and devoid of trapped heat. The gas within the disc attains local thermal equilibrium, resulting in each point on the disc exhibiting a blackbody spectrum. The collective disc manifests a multi-temperature blackbody spectrum, with emissions typically peaking in the soft X-ray band for stellar-mass black holes and in the UV band for supermassive black holes (refer to, e.g.,~\cite{ref70,ref79,Bambi17e, 2021SSRv..217...65B}).}

\textcolor{black}{The thermal spectrum originating from the accretion disc displays a high sensitivity to the position of the inner edge of the disc. Assuming the inner edge aligns with the ISCO radius and possesses independent estimates for the black hole mass, distance, and disc inclination angle, one can employ data fitting to infer the ISCO radius~\cite{Zhang_1997}. It's noteworthy that the ISCO radius relies on the specific background metric. The determination of the ISCO radius can be deduced from the effective potential governing the orbit of a particle around a black hole. Assuming an axially symmetric and stationary spacetime with  metric $g_{\mu \nu}$ expressed in the canonical form, the normalization of the particle 4-velocity $u_\mu u^\mu = -1$ yields the following expression
\begin{eqnarray}
g_{rr} u_r^2 + g_{\theta\theta} u_\theta^2 = V_{\rm eff} ,
\end{eqnarray}
with $V_{\rm eff}$ being the effective potential that reads~\cite{Bambi17e}
\begin{eqnarray}
V_{\rm eff}=\frac{E^2 g_{\phi \phi}+2 E L g_{t \phi}+L^2 g_{tt}}{g_{t\phi}^2-g_{tt}g_{\phi \phi}}-1.
\end{eqnarray}
Here, $E$ and $L$ denote the specific energy and specific angular momentum of the massive particle in an orbit. Expressing these quantities in terms of the metric components, they adopt the following forms
\begin{eqnarray}
E=\frac{-g_{tt}-\Omega g_{t\phi}}{\sqrt{-g_{tt}-2\Omega g_{t\phi}-\Omega^2 g_{\phi \phi}}}\ ,
\end{eqnarray}
\begin{eqnarray}
L=\frac{\Omega g_{\phi \phi}+g_{t \phi}}{\sqrt{-g_{tt}-2\Omega g_{t\phi}-\Omega^2 g_{\phi \phi}}}\ ,
\end{eqnarray}
where $\Omega = d\phi/dt$ represents the angular velocity of the particle, as indicated in~\cite{Bambi17e}.
\beqn
\Omega=\frac{d\phi}{dt} = \frac{-g_{t\phi,r}\pm\sqrt{\{-g_{t\phi,r}\}^2-\{g_{\phi \phi, r}\} \{g_{tt,r}\}}}{g_{\phi \phi, r}},\ 
\eeqn
and $g_{\mu\nu,\rho}$ is defined as $\partial_{\rho}g_{\mu\nu}$. To determine the ISCO radius, it is necessary to solve the set of equations provided below
\begin{eqnarray}\label{isco_eqn}
V_{\rm eff}(r)=0\ ,\quad
V_{\rm eff}'(r)=0\ ,\quad
V_{\rm eff}''(r)=0\ .
\end{eqnarray}
It's noteworthy that these conditions involve the components of the spacetime metric, providing an avenue to restrict the parameters characterizing the spacetime around the central compact object by gauging the ISCO from the continuum spectrum. For instance, if the spacetime metric adheres to the Kerr solution, the estimation of the black hole spin can be achieved~\cite{Zhang_1997}. This methodology is commonly referred to as the Continuum Fitting Method (CFM), extensively employed over the past two decades for spin estimation of stellar-mass black holes~\cite{McClintock:2013vwa}. }

\begin{figure*}[t!]
	\begin{center}
		\includegraphics[width=0.49\linewidth]{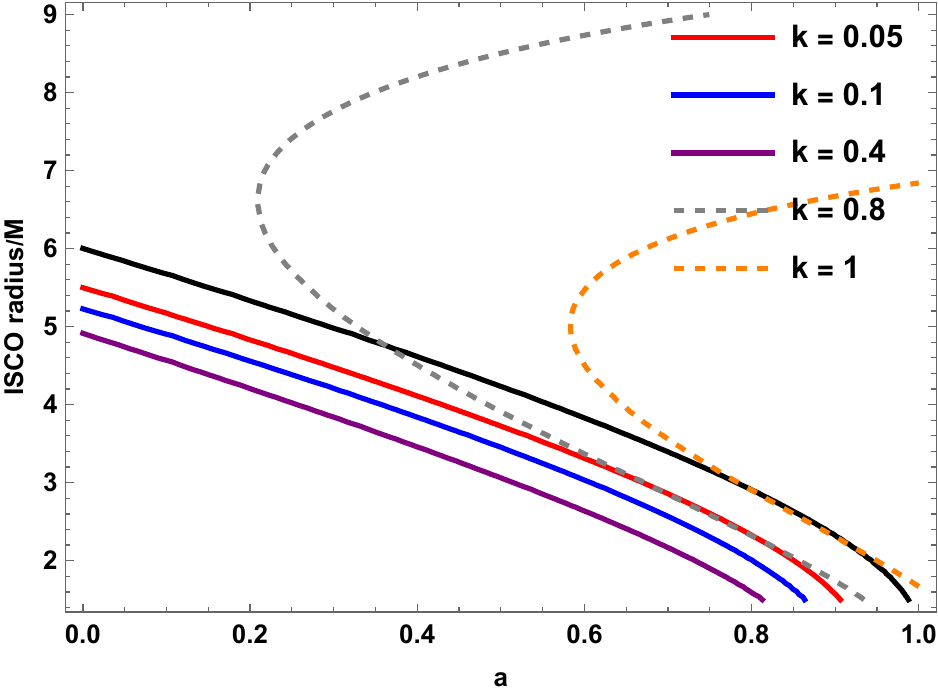}
		\includegraphics[width=0.49\linewidth]{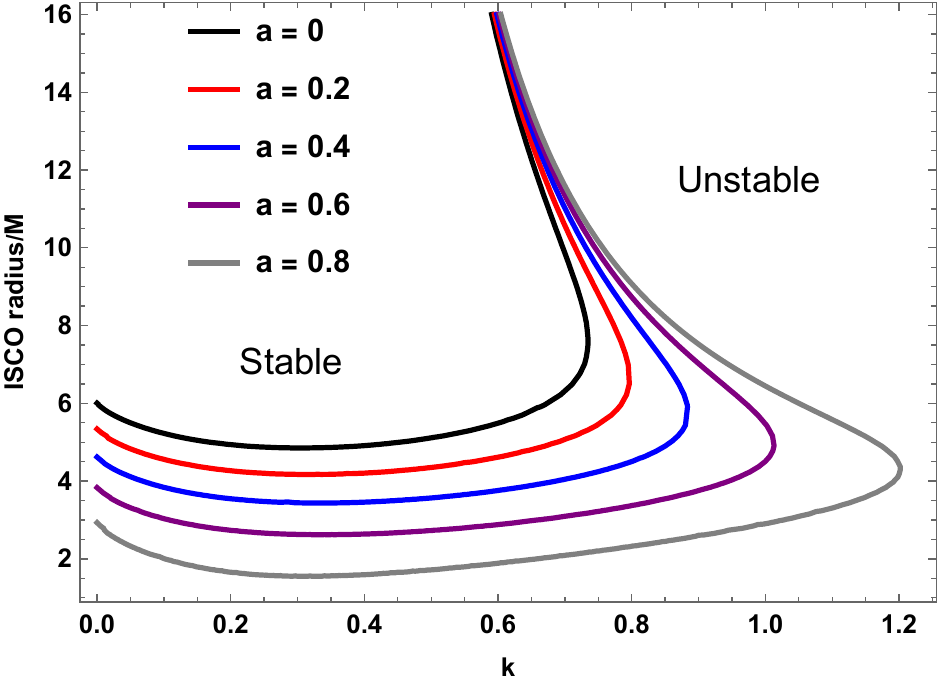}
	\end{center}
	\caption{\textcolor{black}{Influence of the dimensionless spin $a$ and the dimensionless parameter $k$ within the Kerr+PFDM spacetime on the ISCO radius of test particles.}\label{isco}}
\end{figure*}

The relationship between the ISCO radius and the parameters $a$ and $k$ within the Kerr+PFDM spacetime is illustrated in Fig. \ref{isco}. As shown in the left panel, the dependence of the ISCO radius on the spin parameter is not monotonic for all values of PFDM intensity. It is clear that when the PFDM intensity is relatively low, increasing the spin reduces the ISCO radius monotonically (solid lines). However, for higher values of PFDM intensity, there exists a turning point in the lines (dashed lines), making the dependence non-monotonic. This behavior is examined in more detail in the right panel of the figure. Our calculations show that when the PFDM intensity exceeds a certain value, unstable orbits can also be observed. Specifically, if $k \gtrsim 0.5$ , the accretion disk has an inner part with stable orbits and an outer part with no stable orbits. This contradicts the common structure of accretion disks, leading us to restrict the maximum value of PFDM intensity to $k \lesssim 0.5$. This feature may indicate the limitations of using the chosen PFDM model in realistic scenarios. However, later we will demonstrate that even smaller values of $k$ provide satisfactory constraints for spacetime parameters based on observational data.

\begin{figure}[b]
	\begin{center}
		\includegraphics[width=0.95\linewidth]{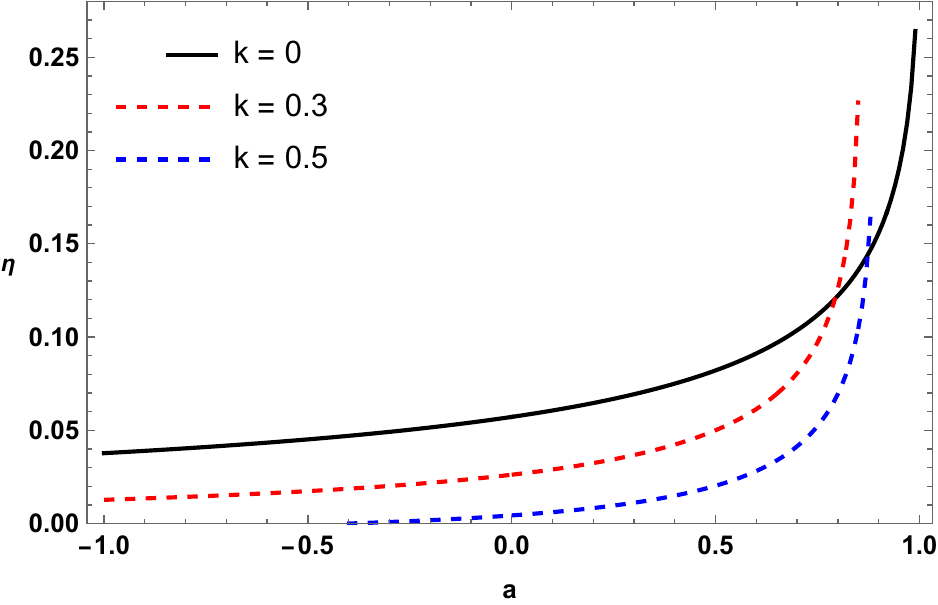}
	\end{center}
	\caption{\textcolor{black}{Change of the radiative efficiency $\eta$ in Kerr+PFDM spacetime with the change of rotation parameter $a$ for distinct values of the parameter $k$.} \label{eff}}
\end{figure}

\textcolor{black}{The radiative efficiency, denoted by $\eta$, characterizing a Novikov-Thorne accretion disc corresponds to the binding energy of a particle in orbit around the black hole at the ISCO radius. This relationship is expressed as follows:
\begin{eqnarray}
\eta=1-E_{\text{isco}}\, . \label{etais}
\end{eqnarray}
Here, $E_{\text{isco}}$ signifies the specific energy of a particle moving at ISCO. Hence, the value of $\eta$ is contingent on the characteristics of the spacetime metric. In the context of the Kerr spacetime, it solely relies on the rotational parameter $a$. However, in the case of the Kerr+PFDM spacetime, $\eta$ is determined by both the values of $a$ and $k$. The graphical representation in Fig.~\ref{eff} illustrates how the radiative efficiency $\eta$ varies for different values of the black hole rotation $a$ and the parameter $k$. When the parameter $k$ is increased, it leads to a reduction in the radiative efficiency $\eta$, whereas an increase in  $a$ results in an augmentation of $\eta$. 
It is worth mentioning that in the Kerr spacetime case ($k=0$), for values of $a$ less than about 0.5 to 0.7, $\eta$ remains relatively flat and nearly constant, with a very small value around 0 to 0.05. However, for values of $a$ greater than about 0.7 to 0.8, $\eta$ increases dramatically. Thus, in one region of the diagram, $\eta$ is nearly independent of $a$ while in another, many different values of $\eta$ correspond to a single value of $a$. This pattern is evident in Fig. 2 from \cite{Piotrovich2023}, which illustrates the same relationship within the Shakura-Sunyaev accretion disk model \cite{Shakura1973}. The plots only show two distinct regions: one where $\eta$ is flat and nearly independent of $a$ (for $-1 < a < 0.75$), and another where $a$ exceeds about 0.75 and $\eta$ can vary significantly between about 0.1 and 1. 
In an approximate sense, black holes featuring a Novikov-Thorne accretion disc with an equivalent radiative efficiency exhibit a similar thermal spectrum \cite{Kong14}. This observation holds the potential for estimating the parameters involved in the spacetime metric.}

\subsection{Relativistic jets }

\textcolor{black}{
Two categories of jets emanate from microquasars~\cite{ref68}:}
\begin{itemize}
    \item \textcolor{black}{Steady, non-relativistic jets, prevalent during the hard-state~\cite{ref83}, persist across a wide range of accretion luminosities.}
    
    \item \textcolor{black}{In contrast, transient or ballistic jets materialize when the source luminosity approaches its Eddington limit during the transition from the hard to soft state. Typically, this type of jet exhibits a relativistic nature and is presumed to originate from near the event horizon~\cite{ref77}. }

\end{itemize}

\begin{figure}
	\begin{center}
		\includegraphics[width= 0.9\linewidth]{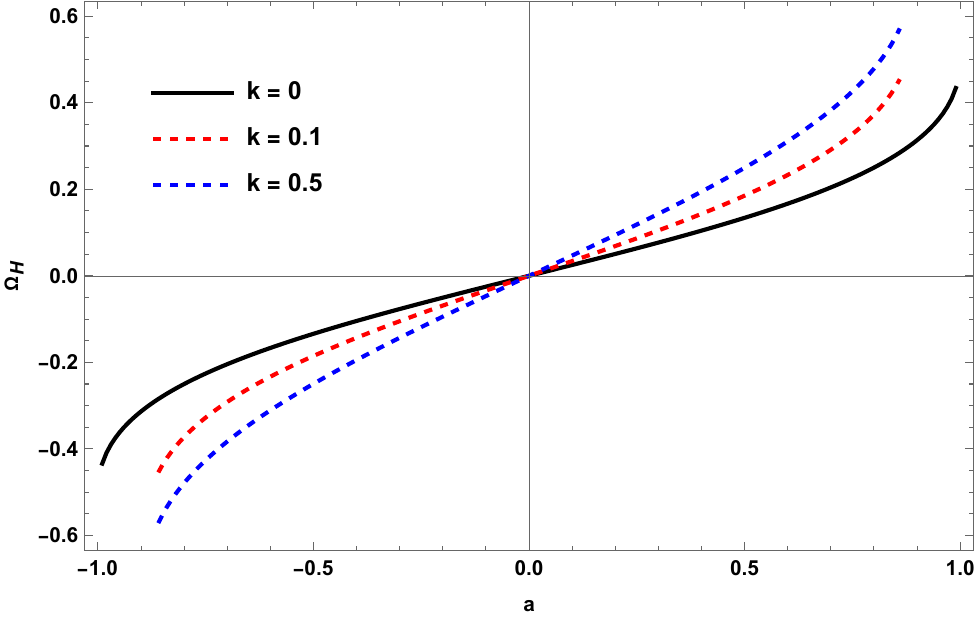}
	\end{center}
	\caption{\textcolor{black}{Change of angular velocity at the horizon with the change of rotation parameter for several values of $k$. \label{W_a}}}
\end{figure}

\textcolor{black}{In this context, we focus on the second category of jets to glean insights into black hole spin and {\bf PFDM intensity}. Despite numerous attempts to elucidate the mechanism behind the generation of relativistic transient jets~\cite{ref85, ref87}, a coherent model to explain observations remains elusive. In this study, we adopt the energy extraction mechanism from a black hole proposed by Blandford and Znajeck~\cite{ref69}. This mechanism, applicable to any axially symmetric spacetime metric, entails the formation of relativistic jets powered by the rotational energy of the black hole through the magnetic field of the current-carrying accretion disc. The total energy-momentum tensor exclusively incorporates the electromagnetic field, while other contributions are disregarded. 
\beqn
T_{\mu \nu}^{tot} \simeq T_{\mu \nu}^{EM}=F_{\mu \alpha} F^\alpha_\nu-\frac{1}{4}g_{\mu \nu} F_{\alpha \beta} F^{\alpha \beta}\ .
\eeqn
In this scenario, the conservation equation is simplified to
\begin{equation}
\nabla^\mu T_{\mu \nu}^{EM}=0\ ,
\end{equation}
where $F_{\mu\nu} = A_{\nu,\mu} - A_{\mu,\nu}$ represents the electromagnetic field tensor associated with the four potential $A_\mu$. In the context of a force-free magnetosphere, the following expression can be readily formulated.
\beqn
\frac{A_{t,r}}{A_\phi,r}=\frac{A_{t,\theta}}{A_{\phi,\theta}}=-\omega(r,\theta)\ .\label{ffree}
\eeqn
Here, $\omega(r,\theta)$ can be construed as an electromagnetic angular velocity~\cite{ref69}. By applying the condition~(\ref{ffree}) to the four potential of the electromagnetic field, $F_{\mu\nu}$ can be expressed subsequently.
\beqn
F_{\mu \nu}=\sqrt{-g} \left(
\begin{array}{cccc}
 0 & - \omega B^{\theta }   &\omega B^r   & 0 \\
 \omega B^{\theta }  & 0 & B^{\phi } & -B^{\theta } \\
 -\omega B^r  & -B^{\phi } & 0 & B^r \\
 0 & B^{\theta } & -B^r & 0 \\
\end{array}
\right)\ .
\eeqn
The expression for the power of relativistic jets in this framework is as follows~\cite{ref69}:
\begin{equation}
P_{BZ}=4 \pi \int_0^{\pi/2} \sqrt{-g} T_t^r d\theta \ , 
\end{equation}
where $T_t^r$ denotes the radial component of the Poynting flux. It is presumed that the jet initiates at the horizon. The radial part of the Poynting flux is defined by:
\beqn
T_t^r=2 r_HM \sin^2\theta (B^r)^2 \omega [\Omega_H-\omega]|_{r=r_H}\, .
\eeqn
The angular velocity $\Omega_H$ is computed at the event horizon $r_H$ and expressed as $$\Omega_H=-\frac{g_{t\phi}}{g_{\phi \phi}}|_{r_H}\, .$$
It's noteworthy to highlight that the initial work by Blandford and Znajek addressed this phenomenon in the slow-rotation limit, applicable when $a$ is close to zero. In this scenario, it was determined that the jet power had to be proportionate to $a^2$~\cite{ref69}. Subsequently, in Ref.~\cite{ref89}, Tchekhovskoy et al. expanded upon the original findings, extending the applicability of the result to nearly the entire range of spin parameters of Kerr black holes. Their conclusion affirmed that, in the Blandford-Znajek model, the leading-order prediction stipulates that the jet power is directly proportional to the square of $\Omega_H$ \cite{ref89}
\beqn
P_{BZ}=\sigma \Phi_{tot}^2 \Omega_H^2\ . \label{PBZ}
\eeqn
Here, $\sigma$ takes on the value of $1/6\pi$ for a split monopole field profile while $\sigma = 0.044$ is for a paraboloidal case \cite{ref69}. The findings in \cite{ref89} were derived within the Kerr metric, and in our study, we assume their validity even when altering the spacetime background. In Eq.~(\ref{PBZ}), $\Phi_{tot}$ represents the magnetic flux and is expressed as: 
\beqn
\Phi_{tot}=2 \pi \int_0^{\pi} \sqrt{-g} |B^r| d\theta\ .
\eeqn
In Fig.~\ref{W_a}, the relationship between the angular velocity $\Omega_H$ and the rotation parameter $a$ is depicted for different intensities of PFDM. The alteration in angular velocity $\Omega_H$ induced by the parameter $k$ significantly influences the power of relativistic jets.}

\section{Constrains from observational data \label{section4.4}}

Given that the radiative efficiency (\ref{etais}) depends on the spacetime metric, its determination can be employed for estimating/constraining the black hole parameters within the corresponding theory of gravity. Subsequently, we will examine certain celestial objects, interpreted in terms of the Kerr+PFDM spacetime: GRS1124-683, H1743-322, GRS1915+105, GROJ1655-40, A0620-00,  and XTEJ1550-564~\cite{ref84,ref90,ref91}.
The spin measurements with the continuum fitting method are obtained when the source is in the soft state and the accretion luminosity is between $\sim5$\% and $\sim30$\% of the Eddington limit. In such conditions, the disk is supposed to be described well by the Novikov-Thorne model, where the inner edge of the disk is at the ISCO radius and outflows and jets are negligible. The accretion luminosity is $L_{acc} = \eta_r \dot{M} c^2$, where $\eta_r = 1 - E_{ISCO}$ and $E_{ISCO}$ is the specific energy if a test-particle at the ISCO radius. For this reason in our model $L_{acc}$ and $\eta$ are determined by the spin value only. In the presence of outflows (not jets!), we would have $L_{acc} = (\eta + \eta_k) \dot{M} c^2$, where $\eta_k$ describes the fraction of energy converted into kinetic energy of the outflow and now $\eta < 1 - E_{ISCO}$.
It is worth mentioning that there are other methods besides the continuum-fitting method that can be used to determine black hole spin in the literature. For example, in \cite{Daly2019, Daly2020} black hole spin values for numerous stellar-mass and supermassive black holes were obtained with so called "Outflow Method". However, results presented using such methods may disagree with the measurements of the continuum-fitting method. For example, for A0620-00 the spin measurement from the continuum-fitting method reported in the literature is $a=0.12\pm0.19$ (see \cite{ref98}) while this source has been found to have very high spin found in \cite{Daly2019}. Since in our paper we assume that the spin measurements from the continuum-fitting method are correct we present the values of spin parameter obtained using this method only as shown in  Table~\ref{Table1}. Notably, the assessments of the black hole rotation parameter $a$ (and the calculated evaluation of the Novikov-Thorne radiative efficiency $\eta$) are all derived under the assumption of the Kerr metric.

\begin{widetext}
\begin{center}
\begin{table}[!ht]
\centering 
    \caption{\textcolor{black}{Characteristics of the binary black hole systems scrutinized in this investigation. The radiative efficiency $\eta$ is derived from the spin measurement utilizing Eq.~(\ref{etais}) within the context of the Kerr metric.}}\label{Table1}
    \begin{tabular}{|c|c|c|c|c|c|}
    \hline
    BH Source & $M \, (M_\odot)$ & $D \, (kpc)$ & $i [deg]$ & $a$ & $\eta$ 
    \\ 
    \hline
  A0620-00 & $6.61 { \pm 0.25}$ & $1.06 { \pm 0.12}$ & $51.0 { \pm 0.9}$ & $0.12 { \pm 0.19}$~\cite{ref98} & $0.061^{+0.009}_{-0.007}$ 
  \\
    \hline
    H1743-322 & $8.0$ & $8.5 {  \pm 0.8}$ & $75.0 { \pm 3.0}$ & $0.2 { \pm 0.3}$ \cite{ref100}& $0.065^{+0.017}_{-0.011}$ 
    \\ 
    \hline
   XTEJ1550-564 & $9.10 { \pm 0.61}$ & $4.38 { \pm 0.5}$ & $74.7 { \pm 3.8}$ & $0.34 { \pm 0.24}$ \cite{steiner2011spin} & $0.072^{+0.017}_{-0.011}$ 
   \\ 
    \hline
   GRS1124-683 & $11.0^{+2.1}_{-1.4}$ & $4.95^{+0.69}_{-0.65}$ & $43.2^{+2.1}_{-2.7}$ & $0.63^{+0.16}_{-0.19}$ \cite{ref109} & $0.095^{+0.025}_{-0.017}$ 
   \\ 
    \hline
   GROJ1655-40 & $6.30 { \pm 0.27}$ & $3.2 { \pm 0.5}$ & $70.2 { \pm 1.9}$ & $0.7 { \pm 0.1}$ \cite{Shafee06}& $0.104^{+0.018}_{-0.013}$ 
   \\ 
    \hline
  GRS1915+105 & $12.4^{+1.7}_{-1.9}$ & $8.6^{+2.0}_{-1.6}$ & $60.0 { \pm 5.0}$ & $a_{*} > 0.98$ \cite{ref118} & $\eta { > 0.234}$ 
  \\ 
    \hline
    \end{tabular}
  \end{table}
  \end{center}
\end{widetext}

\textcolor{black}{In this study, we employ the approach outlined in~\cite{ref84,ref91} to assess the jet power of the six objects under investigation. Conceptually, a bipolar radio jet is envisioned as a symmetrical pair of plasmoids. These plasmoids emit radiation isotropically and possess a slender optical structure, expanding outward from the core source at a relativistic bulk velocity $\beta$. The ratio of observed to emitted flux density for each jet can be expressed as $$S_\nu/S_{\nu,0}=\delta^{3-\alpha} .$$ Here, the Doppler factor is denoted as $\delta$, and the radio spectral index is represented by $\alpha$. The Doppler factor for the brighter jet, approaching the observer, can be straightforwardly expressed in terms of $\beta$, the Lorentz factor $\Gamma$, and the inclination angle $i$ of the jet: $$\delta=(\Gamma [1-\beta \cos i])^{-1} .$$ For the primary emission source, particularly the approaching jet, the observed intensity surpasses the emitted intensity at lower inclinations, while the reverse is true for higher inclinations. In the case of microquasars with mildly relativistic jets, the Doppler boost becomes less than one within the intermediate range of inclinations, approximately between 35 and 55 degrees. We presume that the entire power in the transient jet is proportional to the peak at 5~GHz of the radio flux density (refer to Table~\ref{Table2}). In natural units, the luminosity can be expressed as~\cite{ref84,ref91} 
\begin{eqnarray}
P_{jet}=\left(\frac{\nu}{5~{\rm GHz}}\right) \left(\frac{S_{\nu,0}^{tot}}{\rm Jy}\right) \left(\frac{D}{\rm kpc}\right)^2 \left(\frac{M}{M_{\odot}}\right)^{-1}\, .
\end{eqnarray}
Here, $S_{\nu,0}^{tot}$ represents the beaming associated with the approaching and receding jets~\cite{ref77,ref91}. The Lorentz factor $\Gamma$ linked to the jet is anticipated to fall within the range $2\leq\Gamma\leq5$. The Doppler-corrected jet powers corresponding to Lorentz factors $\Gamma=2$ and $\Gamma=5$ for each source are provided in Table~\ref{Table2}~\cite{ref90,ref94}.}

\begin{table}[!ht]
    \large        
    \caption{Proxy jet power values expressed in units of kpc$^2$~GHz~Jy~$M_{\odot}^{-1}$.}\label{Table2}
    \centering    
    \begin{tabular}{|c|c|c|c|}
    \hline
    BH Source & $(S_{\nu,0})_{max}^{5 GHz}$ (Jy) & $P_{jet}|_{\Gamma=2}$ & $P_{jet}|_{\Gamma=5}$\\     
    \hline
    A0620-00 & 0.203 & 0.13 & 1.6\\
    \hline
    H1743-322 &0.0346 & 7.0 & 140\\
    \hline
     XTEJ1550-564 & 0.265 & 11 & 180\\
    \hline
     GRS1124-683 & 0.45 & 3.9 & 380\\
    \hline
     GROJ1655-40 & 2.42 & 70 & 1600\\
    \hline
     GRS1915+105 & 0.912 & 42 & 660\\
    \hline
    \end{tabular}
  \end{table}

\textcolor{black}{The outcomes presented in Table~\ref{Table2} can be compared with the theoretical projections, contingent on the spacetime metric. According to Eq.\eqref{PBZ}, the jet's power can be articulated as 
\beqn\label{P}
\log P=\log K+2 \log \Omega_H\, .
\eeqn
In this scenario, $K = \sigma \Phi_{tot}^2$. The determination of $K$ involves fitting the observed jet power and $\Omega_H$~\cite{ref84,ref94}. The spacetime metric's influence on the jet power calculations is through the square of the angular velocity of the event horizon $\Omega_H^2$.
In the study in Ref.~\cite{ref94}, the authors derived the best-fitting values for the parameter $K$. They reported $\log K = 2.94 \pm 0.22$ for the Lorentz factor $\Gamma=2$ and $\log K = 4.19 \pm 0.22$ for $\Gamma=5$ (90\% confidence level). Notably, $K$ may not be constant for every source in general. However, it is theorized that the magnetic field strength depends on the mass accretion rate $\Dot{M}$ \cite{ref84, Tchekhovskoyetal.2011}. Since transient jets appear during the transition from the hard to soft state, the Eddington-scaled mass accretion rate remains similar for all the sources. Additionally, the masses of all objects are similar, around 10 Solar masses. For these reasons, one can consider this quantity as constant for the six BH candidates in our list. Assuming $K$ is independent of the spacetime geometry, we then use these determined values of $K$ to constrain the spin parameter and the PFDM intensity parameter of the Kerr+PFDM spacetime based on the observed jet power of the sources in Tab.~\ref{Table2}.}

\subsection{Results}

Below we present our results for every source. 

\begin{itemize}

 \item {\bf Source A0620-00}. 
\textcolor{black}{From the application of the Continuum Fitting Method and assuming the Kerr metric, the estimation for the source's spin parameter is $a=0.12\pm0.19$ at a 68\% confidence level (CL)~\cite{ref98}. This determination of the spin can be translated into a measurement of the radiative efficiency, yielding $\eta = 0.061_{-0.007}^{+0.009}$ as presented in the last column of Tab.~\ref{Table1}.
In accordance with the discussion in Ref.~\cite{Kong14}, the CFM provides an initial approximation of the radiative efficiency for the Novikov-Thorne disc of the source. This value can then be used to impose constraints on the spacetime parameters for a hypothetical non-Kerr black hole. This is the approach taken to constrain the parameters of the Kerr+PFDM spacetime in this study. Fig.~\ref{1Je} illustrates (blue regions) the constraints on the spin parameter $a$ and the parameter $k$ under the condition that the Novikov-Thorne radiative efficiency is $\eta = 0.061_{-0.007}^{+0.009}$. While for $k=0$, we retrieve $a=0.12\pm0.19$, higher values of the spin parameter become permissible for $k > 0$.
The radiative efficiency demonstrates a degeneracy with respect to $a$ and $k$, making it challenging to constrain both parameters without additional measurements.}
\textcolor{black}{The left panel of Fig.~\ref{1Je} corresponds to the case of $\Gamma=2$, while the right panel is for $\Gamma=5$. In both panels, the solid red line represents the central value of $P_{jet}$, and the dashed red lines depict the spacetime parameter values associated with $P_{jet}$ within an error of $0.3 \, dex$ around the central jet power presented in Tab.~\ref{Table2}. The shaded red regions indicate the spacetime parameter values within the error bars.
In panels (a) and (b), it is evident that the Kerr+PFDM spacetime is effective in explaining the jet power of the source. For the case of $\Gamma=2$ in panel (a), the central value of the jet power, which is well explained by the Kerr spacetime with the spin parameter $a\simeq0.05$, can also be suitably explained by the Kerr+PFDM spacetime. The parameters corresponding to the points on the red line account for this agreement. In the right panel (b), representing the Lorentz factor $\Gamma=5$, a slight difference is observed compared to the case of $\Gamma=2$, but it is negligible. It can be observed that an increase in the intensity of dark matter leads to a slight decrease in the value of the spin parameter for a black hole described by the Kerr+PFDM spacetime in order to match the observed jet power values.}
\textcolor{black}{In Fig.~\ref{1Je}, it is evident that the behaviors of the cases with $\Gamma=2$ and $\Gamma=5$ are very similar. The intersection of the blue shaded and red shaded regions theoretically provides values for the spacetime parameters that can simultaneously explain both observations within the error bars using the Kerr+PFDM spacetime. Therefore, from the figure, it can be concluded that the radiative efficiency and jet power of the source A0620-00 can be well explained by the Kerr+PFDM spacetime. The parameters fall within the approximate range $a\simeq0.05\pm0.02$ and $0<k<0.05$, corresponding to the intersections of the blue and red shaded regions.}

\begin{figure*}[t!]
	\begin{center}
		a.
		\includegraphics[width=0.45\linewidth]{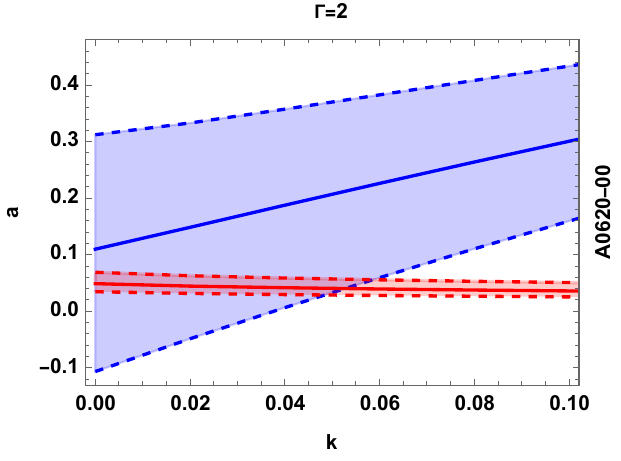}
		b.
		\includegraphics[width=0.45\linewidth]{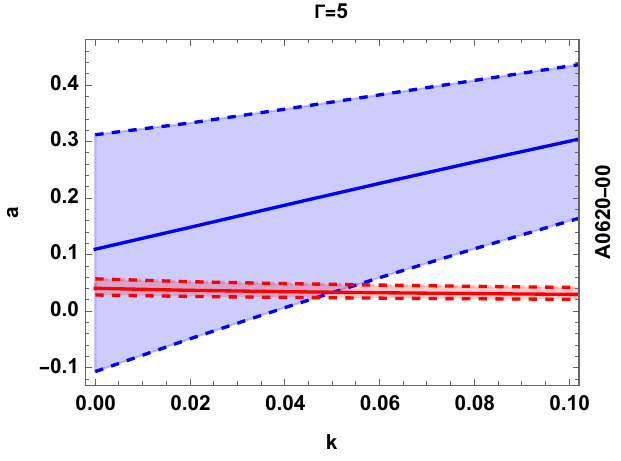}
	\end{center}
	\caption{\textcolor{black}{In the analysis of A0620-00, the blue solid curve in Fig.~\ref{1Je} represents the spin parameter $a$ and the parameter $k$ values that reproduce $\eta = 0.061$ (best-fit value in Tab.~\ref{Table1}). The blue dashed curves correspond to $\eta = 0.054$ and $\eta = 0.070$ (lower and upper constraints in Tab.~\ref{Table1}). The shaded region encompasses all Kerr+PFDM spacetimes that are consistent with the observed thermal spectrum of A0620-00. The red-colored regions indicate the values of spacetime parameters $a$ and $k$ that yield the same jet power as the source. The red solid line corresponds to the central value of the jet power, while the dashed lines represent an error of $0.3 \, dex$. The left panel is for the Lorentz factor $\Gamma=2$, and the right panel is for $\Gamma=5$. The intersection of the blue and red regions satisfies both observational constraints for both $\Gamma=2$ (left panel) and $\Gamma=5$ (right panel). \label{1Je}}}
\end{figure*}

\item {\bf Source H1743-322}. 
\textcolor{black}{The spin of the black hole in H1743-322 has been determined through the CFM, yielding a value of $0.2\pm0.3$ at a 68\% confidence level and $-0.3<a<0.7$ at a 90\% confidence level. Consequently, the radiative efficiency of the source is $\eta = 0.065^{+0.017}_{-0.011}$ at a 68\% confidence level.
Fig.~\ref{2Je} illustrates the constraints on Kerr+PFDM spacetime parameters to account for the observed radiative efficiency of H1743-322. In the case of a Kerr black hole ($k = 0$), the central value of the spin parameter is approximately $a\simeq0.2$. For Kerr+PFDM spacetime with $k=0.2$, the central value of the spin parameter can extend up to $a\simeq0.5$. The dashed blue curves demarcate the boundary of the parameter space permitted by observations.
The red regions in Fig.~\ref{2Je} present the results concerning the jet power of H1743-322. Notably, with an increase in PFDM intensity from $k=0$ to $k=0.2$, the range of the spin parameter required to explain the observed jet power, within the error bars, significantly contracts from $a\simeq0.35\pm0.1$ to $a\simeq0.2\pm0.08$, respectively, for $\Gamma=2$ in the left panel. Conversely, this change is relatively marginal for the case of $\Gamma=5$, as evident from the right panel. Theoretically, the PFDM intensity in the Kerr+PFDM spacetime appears effective in explaining the observed jet power of the source.
Similar to the case of A0620-00, the results for both $\Gamma=2$ and $\Gamma=5$ show almost identical behaviors. The figure highlights that H1743-322 described by the Kerr+PFDM spacetime has favorable parameters ($a\simeq0.3$ and $k\simeq0.055$) that simultaneously explain the central values of both observed radiative efficiency and jet power. The entire red shaded region, with an intensity parameter up to $k\simeq0.11$, proves to be suitable for explaining both observational constraints within the error bars using the Kerr+PFDM spacetime.}

\begin{figure*}[t!]
	\begin{center}
		a.
		\includegraphics[width=0.45\linewidth]{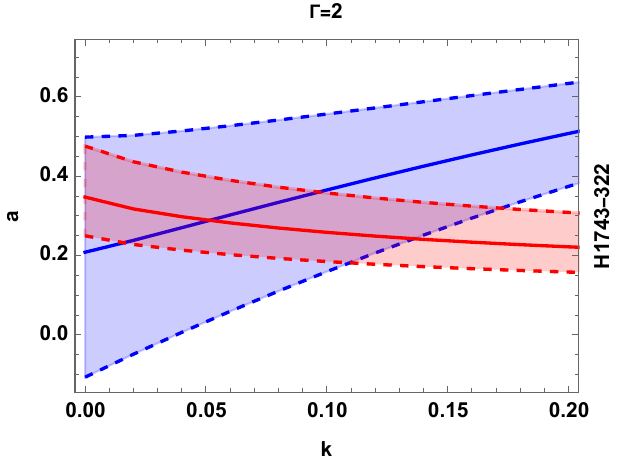}
		b.
		\includegraphics[width=0.45\linewidth]{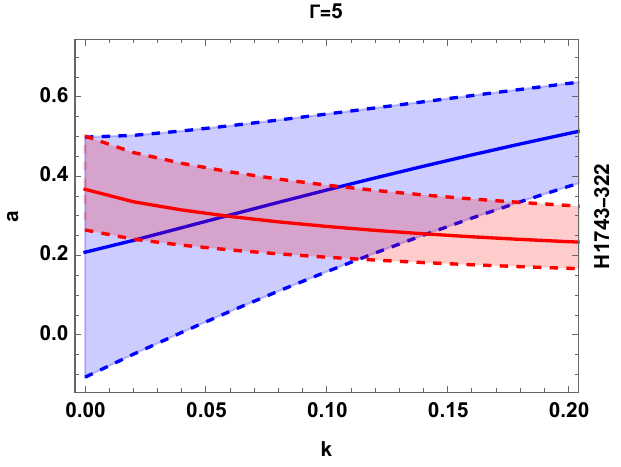}
	\end{center}
	\caption{\textcolor{black}{In the context of H1743-322, the blue shaded region delineates the values of $a$ and $k$ that align with the observed spectrum of the source. Similar to the previous case, Figure (a) corresponds to $\Gamma=2$, and Figure (b) corresponds to $\Gamma=5$, maintaining the same definitions for the lines and regions. From both figures, it is evident that the preferred values for the spin and NUT parameters of the central source are approximately $a\simeq0.3$ and $l\simeq0.055$, respectively.  \label{2Je}}}
\end{figure*}

\item {\bf Source XTE J1550-564}. 
\textcolor{black}{The spin measurement, as reported in \cite{steiner2011spin}, yields values of $a \simeq 0.34\pm 0.24$ at 68\% CL.
Figure \ref{3Je} illustrates the constraints on $a$ and $k$ when the spacetime metric of the black hole is described by the Kerr+PFDM solution. The central value of the spin parameter measurement $a\simeq0.34$ for a Kerr black hole ($k=0$) to explain the radiative efficiency of the source can increase to $a\simeq0.55$ when $k=0.2$.
The values of the parameters $a$ and $k$ describing the observed $P_{jet}$ (red regions) within the error bars are nearly identical in both cases, i.e., for $\Gamma=2$ and $\Gamma=5$. The central value of the spin parameter consistently decreases monotonically with the rise of PFDM intensity, starting from $a\simeq0.42$ when $k=0$ down to $a\simeq0.3$ when $k=0.2$. Consequently, the source XTE J1550-564, initially believed to be a Kerr black hole with the spin parameter $a\simeq0.42\pm0.11$, can also be accommodated by the Kerr+PFDM spacetime. The favorable values of these parameters, describing the observed $\eta$ and $P_{jet}$, correspond to the cross point of the two central blue and red solid lines at $a\simeq0.38$ and $k\simeq0.03$. The figure also indicates that the red shaded region, up to $k\simeq0.1$ in the panels corresponding to the observed jet power, is applicable to explain both observational constraints within the given error bars.}

\begin{figure*}[t!]
	\begin{center}
		a.
		\includegraphics[width=0.45\linewidth]{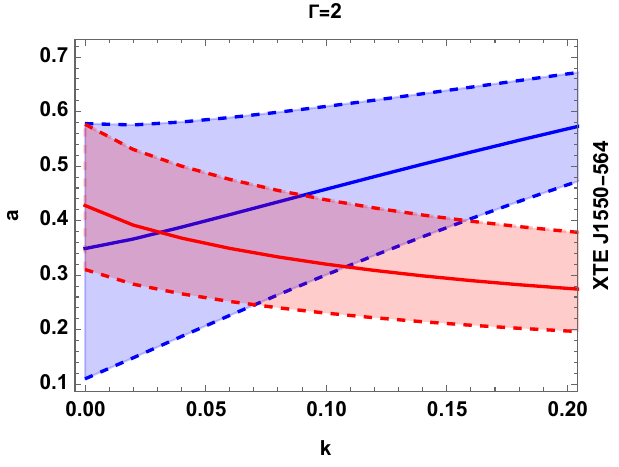}
		b.
		\includegraphics[width=0.45\linewidth]{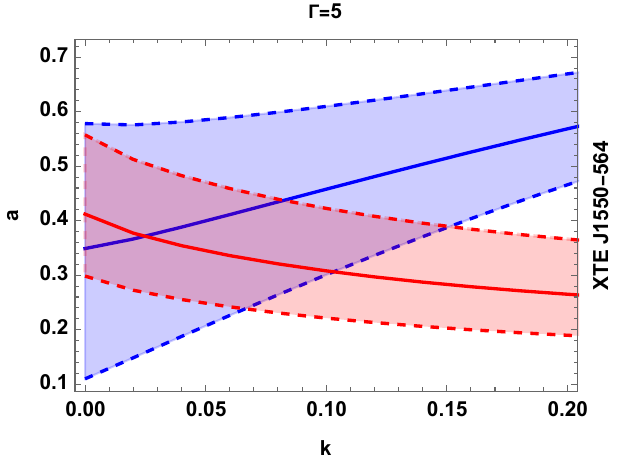}
	\end{center}
	\caption{\textcolor{black}{XTE J1550-564. The interpretation of the diagrams mirrors that of the preceding example; please refer to the corresponding text for an accurate description.\label{3Je}}}
\end{figure*}

\item {\bf Source GRS 1124-683}. 
\textcolor{black}{Utilizing the Continuous Fitting Method (CFM) and assuming the Kerr metric, the spin parameter measurement for the black hole is $0.63^{+0.16}_{-0.19}$ at a 68\% confidence level~\cite{ref109}. Based on this reported measurement, we derive constraints on $a$ and $k$ represented by the blue regions in Fig.~\ref{4Je}. The blue solid line extends to values of the spin and intensity parameters of the Kerr+PFDM metric, with $a\simeq0.7$ and $k\simeq0.2$.
For the observed jet power of the source, indicated by the red-colored regions, notable differences emerge between the cases of Lorentz factor $\Gamma=2$ on the left and $\Gamma=5$ on the right. In the left panel, the spin parameter starts from the range $a\simeq0.26^{+0.11}_{-0.08}$ corresponding to the Kerr case and decreases to $a\simeq0.15^{+0.08}_{-0.03}$ for $k=0.2$. Conversely, in the right panel for $\Gamma=5$, the initial range corresponds to $a\simeq0.57^{+0.17}_{-0.15}$ without the intensity parameter, and it monotonically decreases to $a\simeq0.38^{+0.12}_{-0.1}$ for $k=0.2$.
When seeking to reconcile the two observational constraints, a substantial disparity is evident between the left and right panels. For $\Gamma=2$ in the left panel of Fig.~\ref{4Je}, there is no intersection between the shaded regions of the observed jet power and radiative efficiency. However, in the right panel for $\Gamma=5$, intersecting regions exist, allowing us to identify points within this overlap as capable of explaining both the jet power and radiative efficiency of the source. This suggests that assuming the source is described by Kerr+PFDM spacetime, it corresponds to an object with a relatively high Lorentz factor emitting radio jets.}

\begin{figure*}[t!]
	\begin{center}
		a.
		\includegraphics[width=0.45\linewidth]{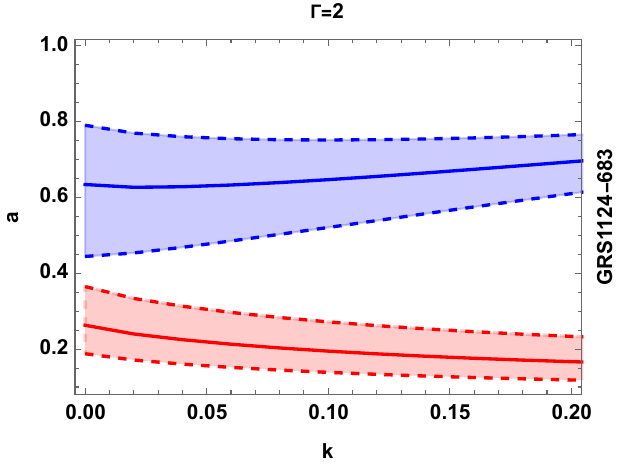}
		b.
		\includegraphics[width=0.45\linewidth]{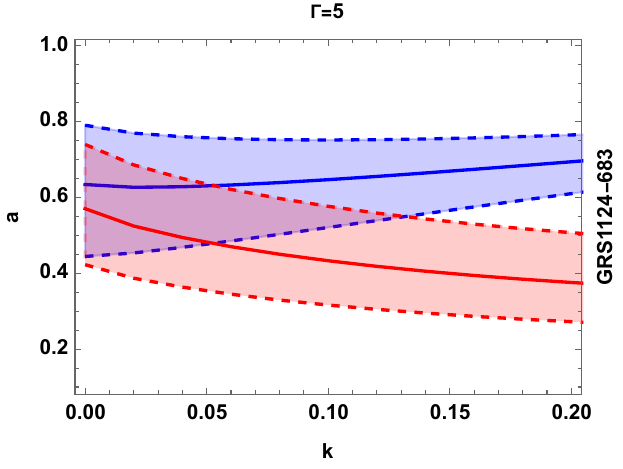}
	\end{center}
	\caption{\textcolor{black}{GRS 1124-683. The left panel indicates that the Kerr+PFDM spacetime, assuming a Lorentz factor $\Gamma=2$, cannot simultaneously account for the observed jet power and radiative efficiency of the source. However, in the right panel, an intersection between the regions becomes apparent for $\Gamma=5$. Further elaboration is provided in the main text. \label{4Je}}}
\end{figure*}

\item {\bf Source GRO J1655-40}. 
\textcolor{black}{The spin measurement obtained from CFM indicates $a\simeq 0.7\pm 0.1$ at a 68\% confidence level~\cite{Shafee06}.
The blue regions in Fig.~\ref{5Je} illustrate the constraints on $a$ and $k$ based on the observed radiative efficiency, assuming the spacetime around the black hole is described by the Kerr+PFDM metric. It is evident that the boundaries of the blue shaded region narrow as the parameter $k$ increases, bringing the error bars closer to the central value of the radiative efficiency.
Red shaded regions in both the left and right panels, representing the observed jet power of the sources, indicate that an increase in the Lorentz factor slightly shifts the regions towards larger values of the spin parameter, without significantly altering the overall behavior of the lines. In general, it can be inferred that an increase in the parameter $k$ from 0 up to 0.4 reduces the central value of the spin parameter from approximately $\sim0.9$ to $\sim0.6$. This suggests that the source GRO J1655-40 might be a fast-rotating black hole when not surrounded by dark matter or a moderately spinning black hole when described by the Kerr+PFDM model.
Discussing the unification of the two constraints, the evaluation for both cases of $\Gamma=2$ and $\Gamma=5$ shows minimal differences in the left and right panels, respectively. Favorable values for the spin and parameter $k$ are found to be $a\simeq0.7$ and $k\simeq0.1$ for Lorentz factor $\Gamma=2$, while for $\Gamma=5$, the spin remains almost the same, but the parameter $k$ increases to $k\simeq0.16$. Theoretically, based on the values of the radiative efficiency presented in Tab.~\ref{Table1} and the jet power in Tab.~\ref{Table2}, the intersection of the regions can be utilized to explain the observational constraints within the given error bars, as depicted in the figure.}

\begin{figure*}[t!]
	\begin{center}
		a.
		\includegraphics[width=0.45\linewidth]{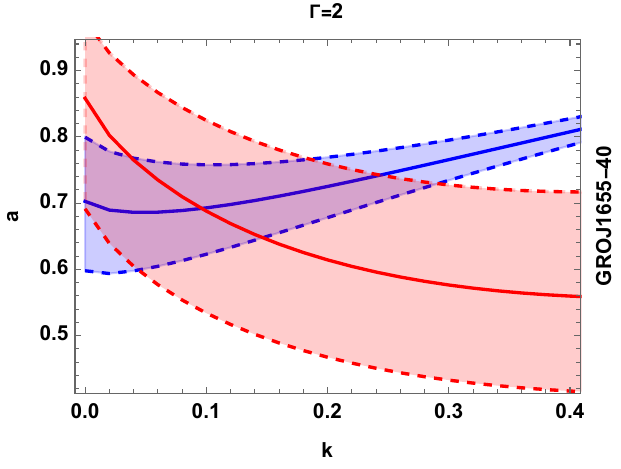}
		b.
		\includegraphics[width=0.45\linewidth]{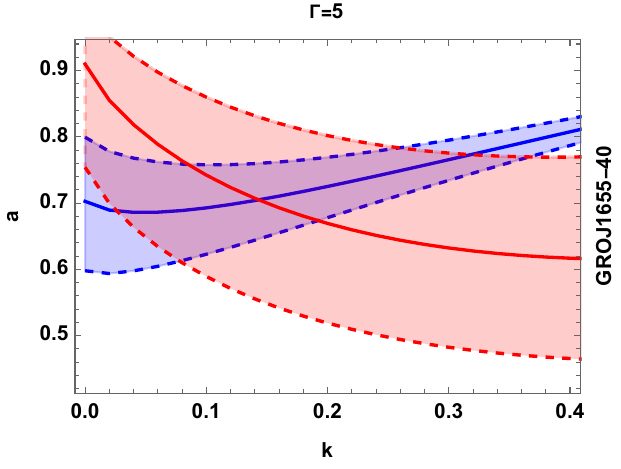}
	\end{center}
	\caption{\textcolor{black}{GRO J1655-40. The left panel corresponds to $\Gamma=2$  and the right panel to $\Gamma=5$. They illustrate that the Kerr+PFDM spacetime is capable of explaining both the observed radiative efficiency, $\eta$, and the jet power, $P_{jet}$. Further discussion on this can be found in the main text.\label{5Je}}}
\end{figure*}

\item {\bf Source GRS~1915+105}. 
\textcolor{black}{In \cite{ref118}, the spin measurement yields $a > 0.98$ at a confidence level of 68\%.
The blue regions displayed in Fig.~\ref{6Je} outline the constraints on $a$ and $k$ concerning the observed radiative efficiency when assuming GRS~1915+105 hosts a Kerr+PFDM black hole. Notably, for this object, the error bars (depicted by blue dashed lines) become less distinguishable from the central value line for relatively higher values of the PFDM intensity.
The trends of the parameters explaining the observed jet power (highlighted in red regions) appear similar for both values of the Lorentz factor. From both the left and right panels, it is evident that the spin parameter needed to describe the observed jet power remains nearly constant starting from around $k\simeq0.3$, irrespective of whether the Lorentz factor is $\Gamma=2$ or $\Gamma=5$.
It's essential to mention that, in the pure Kerr spacetime, the expression \eqref{PBZ} is accurate for spins up to approximately $a\simeq0.95$. For spins exceeding this value, higher-order terms of $\Omega_H$ need consideration \cite{ref89}. However, our analysis, incorporating even higher-order terms ($P_{BZ}=k \Phi_{tot}^2 (\Omega_H^2 + \alpha \Omega_H^4 + \beta \Omega_H^6)$, where $\alpha\simeq 1.38$ and $\beta\simeq -9.2$ according to \cite{ref89}), reveals that corrections from these additional terms are negligible.
From both the left and right panels, it is apparent that there is no overlapping region where the shaded regions of the two observational constraints intersect. This suggests that the observed radiative efficiency and jet power of the source (as given in Tab.~\ref{Table1} and Tab.~\ref{Table2}, respectively) cannot be theoretically explained simultaneously using the Kerr+PFDM spacetime geometry. The figure illustrates that the region capable of explaining the radiative efficiency corresponds to the extreme rotation of the source within the selected range of the parameter $k$.}

\begin{figure*}[t!]
	\begin{center}
		a.
		\includegraphics[width=0.45\linewidth]{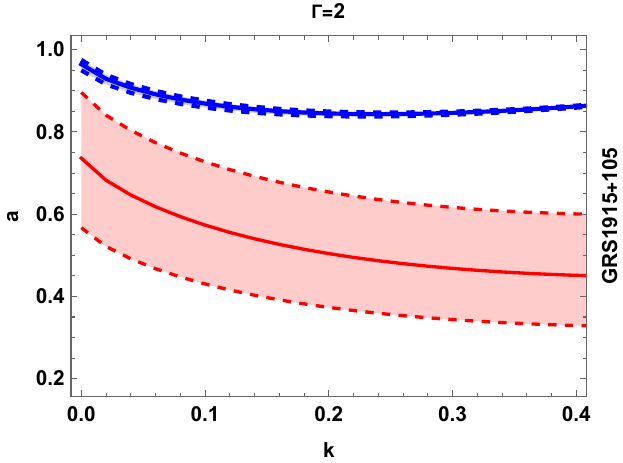}
		b.
		\includegraphics[width=0.45\linewidth]{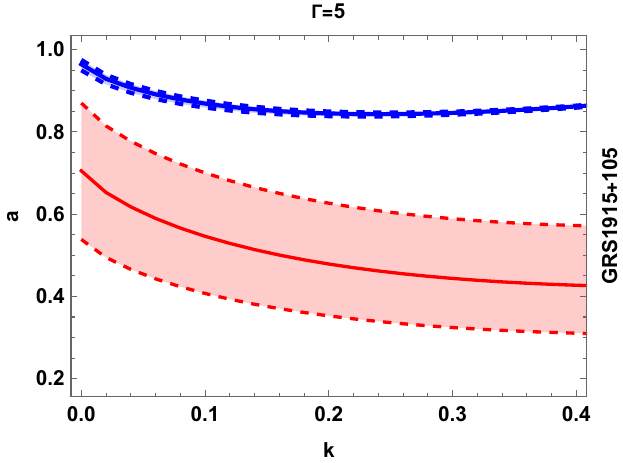}
	\end{center}
	\caption{\textcolor{black}{GRS1915+105. The outcomes presented in the left and right panels indicate that the Kerr+PFDM spacetime encounters challenges in simultaneously explaining both observational constraints. Further discussion is available in the corresponding text.} \label{6Je}}
\end{figure*}

\end{itemize}

The findings in this section suggest that the gravitational sources A0620-00, H1743-322, GRO J1655-40, and XTE J1550-564 have promising prospects to be characterized by the Kerr+PFDM spacetime for both Lorentz factor values, as discussed earlier. For the source GRS1124-683, it is also plausible to elucidate the two observational constraints simultaneously using the Kerr+PFDM spacetime, but a larger Lorentz factor is preferred. Another notable feature of this source is that the observed power jets for $\Gamma=5$ differ from the case of $\Gamma=2$ by a factor of $10^2$, whereas for the other sources, it is on the order of $10^1$. This discrepancy causes the red shaded regions in the graphs to differ significantly, while for the other black hole candidates, this difference is negligible. Conversely, the results for the source GRS1915+105 indicate that the selected spacetime metric may not be suitable for explaining both observational constraints simultaneously. As demonstrated above, there is no intersection of the regions corresponding to these observational constraints.

\section{Conclusion \label{section4.6}}
{In this study, we have employed the Kerr+PFDM metric to characterize the spacetime geometry around well-known sources, namely A0620-00, H1743-322, XTE J1550-564, GRS1124-683, GRO J1655-40, and GRS1915+105. We have provided a brief overview of the spacetime properties, which includes two additional parameters, namely the spin and the parameter $k$ describing the intensity of PFDM (referred to as the ``intensity parameter''). We have demonstrated that, if the spacetime of a source is described by the Kerr+PFDM metric, the observed radiative efficiency of the black hole source can be explained by the corresponding values of these additional spacetime parameters. Our results have indicated that, in all cases, the range of the spin parameter needed to account for the observed radiative efficiency increases with the rise of the parameter $k$.}

{Following this, we employed a similar methodology to assess the jet power of the sources, exploring whether the Kerr+PFDM spacetime could explain the observed jet power within specific ranges of the spacetime parameters $a$ and $k$. Our examinations confirmed that, fundamentally, the Kerr+PFDM spacetime can indeed yield a positive response to this inquiry. We depicted that, for larger values of the parameter $k$, the corresponding range of the spin parameter contracts, in contrast to the trends observed in the case of radiative efficiency. Specifically, for the black hole candidates XTE J1550-564, GRO J1655-40, A0620-00, GRS1915+105, and H1743-322, we observed resemblances in the regions for the values of the Lorentz factor $\Gamma=2$ and $\Gamma=5$ while notable distinctions were identified for the source GRS1124-683. Furthermore, we demonstrated that corresponding to a rapidly rotating black hole in the Kerr spacetime (when $k=0$), the sources may transition to non-rapidly rotating black holes with an increase in the parameter $k$ to accommodate the observed jet power.}

{Subsequently, we combined the outcomes from the previous sections to ascertain whether the Kerr+PFDM spacetime could concurrently elucidate both observed quantities. The findings indicate that the Kerr+PFDM spacetime is indeed suitable for explaining both radiative efficiency and jet power for the sources A0620-00, H1743-322, GRO J1655-40, and XTE J1550-564. However, for the black hole candidate GRS1124-683, the situation appears to shift with changes in the Lorentz factor. In the case of GRS1915+105, the unification of results reveals that the Kerr+PFDM spacetime provides regions for the two observed quantities that do not intersect. This suggests that the chosen spacetime is not conducive to simultaneously reproducing the values of radiative efficiency and jet power as presented in Tab.~\ref{Table1} and Tab.~\ref{Table2}. It is important to note that our results are contingent on the assumptions posited in refs. \cite{ref84} and \cite{ref91}, with some authors challenging these outcomes (see e.g., \cite{Russell2013ws}). Currently, the veracity of either perspective remains uncertain due to the limited number of sources. With an increasing corpus of data and more refined measurements, future analyses could potentially corroborate or refute the correlation proposed by Narayan and McClintock. }

\begin{acknowledgments}
This work was supported by the National Natural Science Foundation of China (NSFC), Grant No.~12261131497 and Grant No.~12250610185, and the Natural Science Foundation of Shanghai, Grant No.~22ZR1403400.
\end{acknowledgments}

\bibliographystyle{apsrev4-1}
\bibliography{gravreferences}

\begin{thebibliography}{55}%
\makeatletter
\providecommand \@ifxundefined [1]{%
 \@ifx{#1\undefined}
}%
\providecommand \@ifnum [1]{%
 \ifnum #1\expandafter \@firstoftwo
 \else \expandafter \@secondoftwo
 \fi
}%
\providecommand \@ifx [1]{%
 \ifx #1\expandafter \@firstoftwo
 \else \expandafter \@secondoftwo
 \fi
}%
\providecommand \natexlab [1]{#1}%
\providecommand \enquote  [1]{``#1''}%
\providecommand \bibnamefont  [1]{#1}%
\providecommand \bibfnamefont [1]{#1}%
\providecommand \citenamefont [1]{#1}%
\providecommand \href@noop [0]{\@secondoftwo}%
\providecommand \href [0]{\begingroup \@sanitize@url \@href}%
\providecommand \@href[1]{\@@startlink{#1}\@@href}%
\providecommand \@@href[1]{\endgroup#1\@@endlink}%
\providecommand \@sanitize@url [0]{\catcode `\\12\catcode `\$12\catcode
  `\&12\catcode `\#12\catcode `\^12\catcode `\_12\catcode `\%12\relax}%
\providecommand \@@startlink[1]{}%
\providecommand \@@endlink[0]{}%
\providecommand \url  [0]{\begingroup\@sanitize@url \@url }%
\providecommand \@url [1]{\endgroup\@href {#1}{\urlprefix }}%
\providecommand \urlprefix  [0]{URL }%
\providecommand \Eprint [0]{\href }%
\providecommand \doibase [0]{http://dx.doi.org/}%
\providecommand \selectlanguage [0]{\@gobble}%
\providecommand \bibinfo  [0]{\@secondoftwo}%
\providecommand \bibfield  [0]{\@secondoftwo}%
\providecommand \translation [1]{[#1]}%
\providecommand \BibitemOpen [0]{}%
\providecommand \bibitemStop [0]{}%
\providecommand \bibitemNoStop [0]{.\EOS\space}%
\providecommand \EOS [0]{\spacefactor3000\relax}%
\providecommand \BibitemShut  [1]{\csname bibitem#1\endcsname}%
\let\auto@bib@innerbib\@empty
\bibitem [{\citenamefont {Fender}\ \emph {et~al.}(2004)\citenamefont {Fender},
  \citenamefont {Belloni},\ and\ \citenamefont {Gallo}}]{ref92}%
  \BibitemOpen
  \bibfield  {author} {\bibinfo {author} {\bibfnamefont {R.~P.}\ \bibnamefont
  {Fender}}, \bibinfo {author} {\bibfnamefont {T.~M.}\ \bibnamefont {Belloni}},
  \ and\ \bibinfo {author} {\bibfnamefont {E.}~\bibnamefont {Gallo}},\ }\href
  {\doibase 10.1111/j.1365-2966.2004.08384.x} {\bibfield  {journal} {\bibinfo
  {journal} {Mon. Not. Roy. Astron. Soc.}\ }\textbf {\bibinfo {volume} {355}},\
  \bibinfo {pages} {1105} (\bibinfo {year} {2004})},\ \Eprint
  {http://arxiv.org/abs/astro-ph/0409360} {arXiv:astro-ph/0409360} \BibitemShut
  {NoStop}%
\bibitem [{\citenamefont {Narayan}\ and\ \citenamefont
  {McClintock}(2012)}]{ref84}%
  \BibitemOpen
  \bibfield  {author} {\bibinfo {author} {\bibfnamefont {R.}~\bibnamefont
  {Narayan}}\ and\ \bibinfo {author} {\bibfnamefont {J.~E.}\ \bibnamefont
  {McClintock}},\ }\href {\doibase 10.1111/j.1745-3933.2011.01181.x} {\bibfield
   {journal} {\bibinfo  {journal} {Mon. Not. Roy. Astron. Soc.}\ }\textbf
  {\bibinfo {volume} {419}},\ \bibinfo {pages} {L69} (\bibinfo {year}
  {2012})},\ \Eprint {http://arxiv.org/abs/1112.0569} {arXiv:1112.0569
  [astro-ph.HE]} \BibitemShut {NoStop}%
\bibitem [{\citenamefont {Bergstr\"om}(2000)}]{Bergstrom:2000pn}%
  \BibitemOpen
  \bibfield  {author} {\bibinfo {author} {\bibfnamefont {L.}~\bibnamefont
  {Bergstr\"om}},\ }\href {\doibase 10.1088/0034-4885/63/5/2r3} {\bibfield
  {journal} {\bibinfo  {journal} {Rept. Prog. Phys.}\ }\textbf {\bibinfo
  {volume} {63}},\ \bibinfo {pages} {793} (\bibinfo {year} {2000})},\ \Eprint
  {http://arxiv.org/abs/hep-ph/0002126} {arXiv:hep-ph/0002126} \BibitemShut
  {NoStop}%
\bibitem [{\citenamefont {Bertone}\ \emph {et~al.}(2005)\citenamefont
  {Bertone}, \citenamefont {Hooper},\ and\ \citenamefont
  {Silk}}]{Bertone:2004pz}%
  \BibitemOpen
  \bibfield  {author} {\bibinfo {author} {\bibfnamefont {G.}~\bibnamefont
  {Bertone}}, \bibinfo {author} {\bibfnamefont {D.}~\bibnamefont {Hooper}}, \
  and\ \bibinfo {author} {\bibfnamefont {J.}~\bibnamefont {Silk}},\ }\href
  {\doibase 10.1016/j.physrep.2004.08.031} {\bibfield  {journal} {\bibinfo
  {journal} {Phys. Rept.}\ }\textbf {\bibinfo {volume} {405}},\ \bibinfo
  {pages} {279} (\bibinfo {year} {2005})},\ \Eprint
  {http://arxiv.org/abs/hep-ph/0404175} {arXiv:hep-ph/0404175} \BibitemShut
  {NoStop}%
\bibitem [{\citenamefont {Feng}(2010)}]{Feng:2010gw}%
  \BibitemOpen
  \bibfield  {author} {\bibinfo {author} {\bibfnamefont {J.~L.}\ \bibnamefont
  {Feng}},\ }\href {\doibase 10.1146/annurev-astro-082708-101659} {\bibfield
  {journal} {\bibinfo  {journal} {Ann. Rev. Astron. Astrophys.}\ }\textbf
  {\bibinfo {volume} {48}},\ \bibinfo {pages} {495} (\bibinfo {year} {2010})},\
  \Eprint {http://arxiv.org/abs/1003.0904} {arXiv:1003.0904 [astro-ph.CO]}
  \BibitemShut {NoStop}%
\bibitem [{\citenamefont {Chan}\ and\ \citenamefont {Lee}(2023)}]{Chan2022gqd}%
  \BibitemOpen
  \bibfield  {author} {\bibinfo {author} {\bibfnamefont {M.~H.}\ \bibnamefont
  {Chan}}\ and\ \bibinfo {author} {\bibfnamefont {C.~M.}\ \bibnamefont {Lee}},\
  }\href {\doibase 10.3847/2041-8213/acaafa} {\bibfield  {journal} {\bibinfo
  {journal} {Astrophys. J. Lett.}\ }\textbf {\bibinfo {volume} {943}},\
  \bibinfo {pages} {L11} (\bibinfo {year} {2023})},\ \Eprint
  {http://arxiv.org/abs/2212.05664} {arXiv:2212.05664 [astro-ph.HE]}
  \BibitemShut {NoStop}%
\bibitem [{\citenamefont {Steiner}\ \emph {et~al.}(2013)\citenamefont
  {Steiner}, \citenamefont {McClintock},\ and\ \citenamefont
  {Narayan}}]{ref91}%
  \BibitemOpen
  \bibfield  {author} {\bibinfo {author} {\bibfnamefont {J.~F.}\ \bibnamefont
  {Steiner}}, \bibinfo {author} {\bibfnamefont {J.~E.}\ \bibnamefont
  {McClintock}}, \ and\ \bibinfo {author} {\bibfnamefont {R.}~\bibnamefont
  {Narayan}},\ }\href {\doibase 10.1088/0004-637X/762/2/104} {\bibfield
  {journal} {\bibinfo  {journal} {Astrophys. J.}\ }\textbf {\bibinfo {volume}
  {762}},\ \bibinfo {pages} {104} (\bibinfo {year} {2013})},\ \Eprint
  {http://arxiv.org/abs/1211.5379} {arXiv:1211.5379 [astro-ph.HE]} \BibitemShut
  {NoStop}%
\bibitem [{\citenamefont {Bambi}(2012{\natexlab{a}})}]{ref78}%
  \BibitemOpen
  \bibfield  {author} {\bibinfo {author} {\bibfnamefont {C.}~\bibnamefont
  {Bambi}},\ }\href {\doibase 10.1103/PhysRevD.85.043002} {\bibfield  {journal}
  {\bibinfo  {journal} {Phys. Rev. D}\ }\textbf {\bibinfo {volume} {85}},\
  \bibinfo {pages} {043002} (\bibinfo {year} {2012}{\natexlab{a}})},\ \Eprint
  {http://arxiv.org/abs/1201.1638} {arXiv:1201.1638 [gr-qc]} \BibitemShut
  {NoStop}%
\bibitem [{\citenamefont {Banerjee}\ \emph {et~al.}(2021)\citenamefont
  {Banerjee}, \citenamefont {Mandal},\ and\ \citenamefont
  {SenGupta}}]{Banerjee2020ubc}%
  \BibitemOpen
  \bibfield  {author} {\bibinfo {author} {\bibfnamefont {I.}~\bibnamefont
  {Banerjee}}, \bibinfo {author} {\bibfnamefont {B.}~\bibnamefont {Mandal}}, \
  and\ \bibinfo {author} {\bibfnamefont {S.}~\bibnamefont {SenGupta}},\ }\href
  {\doibase 10.1103/PhysRevD.103.044046} {\bibfield  {journal} {\bibinfo
  {journal} {Phys. Rev. D}\ }\textbf {\bibinfo {volume} {103}},\ \bibinfo
  {pages} {044046} (\bibinfo {year} {2021})},\ \Eprint
  {http://arxiv.org/abs/2007.03947} {arXiv:2007.03947 [gr-qc]} \BibitemShut
  {NoStop}%
\bibitem [{\citenamefont {Narzilloev}\ \emph {et~al.}(2023)\citenamefont
  {Narzilloev}, \citenamefont {Abdujabbarov}, \citenamefont {Ahmedov},\ and\
  \citenamefont {Bambi}}]{Narzilloev2023h}%
  \BibitemOpen
  \bibfield  {author} {\bibinfo {author} {\bibfnamefont {B.}~\bibnamefont
  {Narzilloev}}, \bibinfo {author} {\bibfnamefont {A.}~\bibnamefont
  {Abdujabbarov}}, \bibinfo {author} {\bibfnamefont {B.}~\bibnamefont
  {Ahmedov}}, \ and\ \bibinfo {author} {\bibfnamefont {C.}~\bibnamefont
  {Bambi}},\ }\href {\doibase 10.1103/PhysRevD.108.103013} {\bibfield
  {journal} {\bibinfo  {journal} {Phys. Rev. D}\ }\textbf {\bibinfo {volume}
  {108}},\ \bibinfo {pages} {103013} (\bibinfo {year} {2023})}\BibitemShut
  {NoStop}%
\bibitem [{\citenamefont {Bambi}(2012{\natexlab{b}})}]{PhysRevD86123013}%
  \BibitemOpen
  \bibfield  {author} {\bibinfo {author} {\bibfnamefont {C.}~\bibnamefont
  {Bambi}},\ }\href {\doibase 10.1103/PhysRevD.86.123013} {\bibfield  {journal}
  {\bibinfo  {journal} {Phys. Rev. D}\ }\textbf {\bibinfo {volume} {86}},\
  \bibinfo {pages} {123013} (\bibinfo {year} {2012}{\natexlab{b}})}\BibitemShut
  {NoStop}%
\bibitem [{\citenamefont {{Hakimov}}\ \emph {et~al.}(2017)\citenamefont
  {{Hakimov}}, \citenamefont {{Abdujabbarov}},\ and\ \citenamefont
  {{Narzilloev}}}]{Hakimov17}%
  \BibitemOpen
  \bibfield  {author} {\bibinfo {author} {\bibfnamefont {A.}~\bibnamefont
  {{Hakimov}}}, \bibinfo {author} {\bibfnamefont {A.}~\bibnamefont
  {{Abdujabbarov}}}, \ and\ \bibinfo {author} {\bibfnamefont {B.}~\bibnamefont
  {{Narzilloev}}},\ }\href {\doibase 10.1142/S0217751X17501160} {\bibfield
  {journal} {\bibinfo  {journal} {International Journal of Modern Physics A}\
  }\textbf {\bibinfo {volume} {32}},\ \bibinfo {eid} {1750116} (\bibinfo {year}
  {2017})}\BibitemShut {NoStop}%
\bibitem [{\citenamefont {Rayimbaev}\ \emph {et~al.}(2021)\citenamefont
  {Rayimbaev}, \citenamefont {Narzilloev}, \citenamefont {Abdujabbarov},\ and\
  \citenamefont {Ahmedov}}]{Narzilloev21c}%
  \BibitemOpen
  \bibfield  {author} {\bibinfo {author} {\bibfnamefont {J.}~\bibnamefont
  {Rayimbaev}}, \bibinfo {author} {\bibfnamefont {B.}~\bibnamefont
  {Narzilloev}}, \bibinfo {author} {\bibfnamefont {A.}~\bibnamefont
  {Abdujabbarov}}, \ and\ \bibinfo {author} {\bibfnamefont {B.}~\bibnamefont
  {Ahmedov}},\ }\href {\doibase 10.3390/galaxies9040071} {\bibfield  {journal}
  {\bibinfo  {journal} {Galaxies}\ }\textbf {\bibinfo {volume} {9}} (\bibinfo
  {year} {2021}),\ 10.3390/galaxies9040071}\BibitemShut {NoStop}%
\bibitem [{\citenamefont {Narzilloev}\ \emph {et~al.}(2021)\citenamefont
  {Narzilloev}, \citenamefont {Rayimbaev}, \citenamefont {Abdujabbarov},\ and\
  \citenamefont {Ahmedov}}]{Narzilloev21d}%
  \BibitemOpen
  \bibfield  {author} {\bibinfo {author} {\bibfnamefont {B.}~\bibnamefont
  {Narzilloev}}, \bibinfo {author} {\bibfnamefont {J.}~\bibnamefont
  {Rayimbaev}}, \bibinfo {author} {\bibfnamefont {A.}~\bibnamefont
  {Abdujabbarov}}, \ and\ \bibinfo {author} {\bibfnamefont {B.}~\bibnamefont
  {Ahmedov}},\ }\href {\doibase 10.3390/galaxies9030063} {\bibfield  {journal}
  {\bibinfo  {journal} {Galaxies}\ }\textbf {\bibinfo {volume} {9}} (\bibinfo
  {year} {2021}),\ 10.3390/galaxies9030063}\BibitemShut {NoStop}%
\bibitem [{\citenamefont {Narzilloev}\ and\ \citenamefont
  {Ahmedov}(2022)}]{Narzilloev22a}%
  \BibitemOpen
  \bibfield  {author} {\bibinfo {author} {\bibfnamefont {B.}~\bibnamefont
  {Narzilloev}}\ and\ \bibinfo {author} {\bibfnamefont {B.}~\bibnamefont
  {Ahmedov}},\ }\href {\doibase 10.3390/sym14091765} {\bibfield  {journal}
  {\bibinfo  {journal} {Symmetry}\ }\textbf {\bibinfo {volume} {14}} (\bibinfo
  {year} {2022}),\ 10.3390/sym14091765}\BibitemShut {NoStop}%
\bibitem [{\citenamefont {{Narzilloev}}\ and\ \citenamefont
  {{Ahmedov}}(2023)}]{Narzilloev23}%
  \BibitemOpen
  \bibfield  {author} {\bibinfo {author} {\bibfnamefont {B.}~\bibnamefont
  {{Narzilloev}}}\ and\ \bibinfo {author} {\bibfnamefont {B.}~\bibnamefont
  {{Ahmedov}}},\ }\href {\doibase 10.3390/sym15020293} {\bibfield  {journal}
  {\bibinfo  {journal} {Symmetry}\ }\textbf {\bibinfo {volume} {15}},\ \bibinfo
  {pages} {293} (\bibinfo {year} {2023})}\BibitemShut {NoStop}%
\bibitem [{\citenamefont {Narzilloev}\ \emph {et~al.}(2022)\citenamefont
  {Narzilloev}, \citenamefont {Abdujabbarov},\ and\ \citenamefont
  {Hakimov}}]{Narzilloev22c}%
  \BibitemOpen
  \bibfield  {author} {\bibinfo {author} {\bibfnamefont {B.}~\bibnamefont
  {Narzilloev}}, \bibinfo {author} {\bibfnamefont {A.}~\bibnamefont
  {Abdujabbarov}}, \ and\ \bibinfo {author} {\bibfnamefont {A.}~\bibnamefont
  {Hakimov}},\ }\href {\doibase 10.1142/S0217751X22501445} {\bibfield
  {journal} {\bibinfo  {journal} {International Journal of Modern Physics A}\
  }\textbf {\bibinfo {volume} {37}},\ \bibinfo {pages} {2250144} (\bibinfo
  {year} {2022})},\ \Eprint
  {http://arxiv.org/abs/https://doi.org/10.1142/S0217751X22501445}
  {https://doi.org/10.1142/S0217751X22501445} \BibitemShut {NoStop}%
\bibitem [{\citenamefont {{Mirzaev}}\ \emph {et~al.}(2023)\citenamefont
  {{Mirzaev}}, \citenamefont {{Li}}, \citenamefont {{Narzilloev}},
  \citenamefont {{Hussain}}, \citenamefont {{Abdujabbarov}},\ and\
  \citenamefont {{Ahmedov}}}]{Narzilloev23a}%
  \BibitemOpen
  \bibfield  {author} {\bibinfo {author} {\bibfnamefont {T.}~\bibnamefont
  {{Mirzaev}}}, \bibinfo {author} {\bibfnamefont {S.}~\bibnamefont {{Li}}},
  \bibinfo {author} {\bibfnamefont {B.}~\bibnamefont {{Narzilloev}}}, \bibinfo
  {author} {\bibfnamefont {I.}~\bibnamefont {{Hussain}}}, \bibinfo {author}
  {\bibfnamefont {A.}~\bibnamefont {{Abdujabbarov}}}, \ and\ \bibinfo {author}
  {\bibfnamefont {B.}~\bibnamefont {{Ahmedov}}},\ }\href {\doibase
  10.1140/epjp/s13360-022-03632-4} {\bibfield  {journal} {\bibinfo  {journal}
  {European Physical Journal Plus}\ }\textbf {\bibinfo {volume} {138}},\
  \bibinfo {eid} {47} (\bibinfo {year} {2023})}\BibitemShut {NoStop}%
\bibitem [{\citenamefont {Narzilloev}\ and\ \citenamefont
  {Ahmedov}(2023{\natexlab{a}})}]{Narzilloev2023b}%
  \BibitemOpen
  \bibfield  {author} {\bibinfo {author} {\bibfnamefont {B.}~\bibnamefont
  {Narzilloev}}\ and\ \bibinfo {author} {\bibfnamefont {B.}~\bibnamefont
  {Ahmedov}},\ }\href {\doibase 10.1142/S0217751X23500264} {\bibfield
  {journal} {\bibinfo  {journal} {Int. J. Mod. Phys. A}\ }\textbf {\bibinfo
  {volume} {38}},\ \bibinfo {pages} {2350026} (\bibinfo {year}
  {2023}{\natexlab{a}})}\BibitemShut {NoStop}%
\bibitem [{\citenamefont {{Abdulxamidov, Farrux}}\ \emph
  {et~al.}(2023)\citenamefont {{Abdulxamidov, Farrux}}, \citenamefont
  {{Benavides-Gallego, Carlos A.}}, \citenamefont {{Narzilloev, Bakhtiyor}},
  \citenamefont {{Hussain, Ibrar}}, \citenamefont {{Abdujabbarov, Ahmadjon}},
  \citenamefont {{Ahmedov, Bobomurat}},\ and\ \citenamefont {{Xu,
  Haiguang}}}]{Narzilloev2023c}%
  \BibitemOpen
  \bibfield  {author} {\bibinfo {author} {\bibnamefont {{Abdulxamidov,
  Farrux}}}, \bibinfo {author} {\bibnamefont {{Benavides-Gallego, Carlos A.}}},
  \bibinfo {author} {\bibnamefont {{Narzilloev, Bakhtiyor}}}, \bibinfo {author}
  {\bibnamefont {{Hussain, Ibrar}}}, \bibinfo {author} {\bibnamefont
  {{Abdujabbarov, Ahmadjon}}}, \bibinfo {author} {\bibnamefont {{Ahmedov,
  Bobomurat}}}, \ and\ \bibinfo {author} {\bibnamefont {{Xu, Haiguang}}},\
  }\href {\doibase 10.1140/epjp/s13360-023-04283-9} {\bibfield  {journal}
  {\bibinfo  {journal} {Eur. Phys. J. Plus}\ }\textbf {\bibinfo {volume}
  {138}},\ \bibinfo {pages} {635} (\bibinfo {year} {2023})}\BibitemShut
  {NoStop}%
\bibitem [{\citenamefont {Alibekov}\ \emph {et~al.}(2023)\citenamefont
  {Alibekov}, \citenamefont {Narzilloev}, \citenamefont {Abdujabbarov},\ and\
  \citenamefont {Ahmedov}}]{Narzilloev2023d}%
  \BibitemOpen
  \bibfield  {author} {\bibinfo {author} {\bibfnamefont {H.}~\bibnamefont
  {Alibekov}}, \bibinfo {author} {\bibfnamefont {B.}~\bibnamefont
  {Narzilloev}}, \bibinfo {author} {\bibfnamefont {A.}~\bibnamefont
  {Abdujabbarov}}, \ and\ \bibinfo {author} {\bibfnamefont {B.}~\bibnamefont
  {Ahmedov}},\ }\href {\doibase 10.3390/sym15071414} {\bibfield  {journal}
  {\bibinfo  {journal} {Symmetry}\ }\textbf {\bibinfo {volume} {15}} (\bibinfo
  {year} {2023}),\ 10.3390/sym15071414}\BibitemShut {NoStop}%
\bibitem [{\citenamefont {Alloqulov}\ \emph {et~al.}(2023)\citenamefont
  {Alloqulov}, \citenamefont {Narzilloev}, \citenamefont {Hussain},
  \citenamefont {Abdujabbarov},\ and\ \citenamefont
  {Ahmedov}}]{Narzilloev2023f}%
  \BibitemOpen
  \bibfield  {author} {\bibinfo {author} {\bibfnamefont {M.}~\bibnamefont
  {Alloqulov}}, \bibinfo {author} {\bibfnamefont {B.}~\bibnamefont
  {Narzilloev}}, \bibinfo {author} {\bibfnamefont {I.}~\bibnamefont {Hussain}},
  \bibinfo {author} {\bibfnamefont {A.}~\bibnamefont {Abdujabbarov}}, \ and\
  \bibinfo {author} {\bibfnamefont {B.}~\bibnamefont {Ahmedov}},\ }\href
  {\doibase 10.1016/j.cjph.2023.07.005} {\bibfield  {journal} {\bibinfo
  {journal} {Chin. J. Phys.}\ }\textbf {\bibinfo {volume} {85}},\ \bibinfo
  {pages} {302} (\bibinfo {year} {2023})}\BibitemShut {NoStop}%
\bibitem [{\citenamefont {Narzilloev}\ and\ \citenamefont
  {Ahmedov}(2023{\natexlab{b}})}]{Narzilloev2023g}%
  \BibitemOpen
  \bibfield  {author} {\bibinfo {author} {\bibfnamefont {B.}~\bibnamefont
  {Narzilloev}}\ and\ \bibinfo {author} {\bibfnamefont {B.}~\bibnamefont
  {Ahmedov}},\ }\href {\doibase 10.1142/S0218271823500645} {\bibfield
  {journal} {\bibinfo  {journal} {Int. J. Mod. Phys. D}\ }\textbf {\bibinfo
  {volume} {32}},\ \bibinfo {pages} {2350064} (\bibinfo {year}
  {2023}{\natexlab{b}})}\BibitemShut {NoStop}%
\bibitem [{\citenamefont {Li}\ and\ \citenamefont {Yang}(2012)}]{Li2012zx}%
  \BibitemOpen
  \bibfield  {author} {\bibinfo {author} {\bibfnamefont {M.-H.}\ \bibnamefont
  {Li}}\ and\ \bibinfo {author} {\bibfnamefont {K.-C.}\ \bibnamefont {Yang}},\
  }\href {\doibase 10.1103/PhysRevD.86.123015} {\bibfield  {journal} {\bibinfo
  {journal} {Phys. Rev. D}\ }\textbf {\bibinfo {volume} {86}},\ \bibinfo
  {pages} {123015} (\bibinfo {year} {2012})},\ \Eprint
  {http://arxiv.org/abs/1204.3178} {arXiv:1204.3178 [astro-ph.CO]} \BibitemShut
  {NoStop}%
\bibitem [{\citenamefont {Liang}\ \emph {et~al.}(2023)\citenamefont {Liang},
  \citenamefont {Hu}, \citenamefont {Wu},\ and\ \citenamefont
  {An}}]{Liang2023jrj}%
  \BibitemOpen
  \bibfield  {author} {\bibinfo {author} {\bibfnamefont {X.}~\bibnamefont
  {Liang}}, \bibinfo {author} {\bibfnamefont {Y.-P.}\ \bibnamefont {Hu}},
  \bibinfo {author} {\bibfnamefont {C.-H.}\ \bibnamefont {Wu}}, \ and\ \bibinfo
  {author} {\bibfnamefont {Y.-S.}\ \bibnamefont {An}},\ }\href {\doibase
  10.1140/epjc/s10052-023-12200-8} {\bibfield  {journal} {\bibinfo  {journal}
  {Eur. Phys. J. C}\ }\textbf {\bibinfo {volume} {83}},\ \bibinfo {pages}
  {1009} (\bibinfo {year} {2023})},\ \Eprint {http://arxiv.org/abs/2308.00308}
  {arXiv:2308.00308 [gr-qc]} \BibitemShut {NoStop}%
\bibitem [{\citenamefont {Bambi}(2017)}]{Bambi17e}%
  \BibitemOpen
  \bibfield  {author} {\bibinfo {author} {\bibfnamefont {C.}~\bibnamefont
  {Bambi}},\ }\href@noop {} {\emph {\bibinfo {title} {Black Holes: A Laboratory
  for Testing Strong Gravity}}}\ (\bibinfo  {publisher} {Springer, Singapore},\
  \bibinfo {year} {2017})\BibitemShut {NoStop}%
\bibitem [{\citenamefont {Bovy}\ and\ \citenamefont
  {Tremaine}(2012)}]{Bovy2012tw}%
  \BibitemOpen
  \bibfield  {author} {\bibinfo {author} {\bibfnamefont {J.}~\bibnamefont
  {Bovy}}\ and\ \bibinfo {author} {\bibfnamefont {S.}~\bibnamefont
  {Tremaine}},\ }\href {\doibase 10.1088/0004-637X/756/1/89} {\bibfield
  {journal} {\bibinfo  {journal} {Astrophys. J.}\ }\textbf {\bibinfo {volume}
  {756}},\ \bibinfo {pages} {89} (\bibinfo {year} {2012})},\ \Eprint
  {http://arxiv.org/abs/1205.4033} {arXiv:1205.4033 [astro-ph.GA]} \BibitemShut
  {NoStop}%
\bibitem [{\citenamefont {Xu}\ \emph {et~al.}(2018)\citenamefont {Xu},
  \citenamefont {Wang},\ and\ \citenamefont {Hou}}]{Xu2017bpz}%
  \BibitemOpen
  \bibfield  {author} {\bibinfo {author} {\bibfnamefont {Z.}~\bibnamefont
  {Xu}}, \bibinfo {author} {\bibfnamefont {J.}~\bibnamefont {Wang}}, \ and\
  \bibinfo {author} {\bibfnamefont {X.}~\bibnamefont {Hou}},\ }\href {\doibase
  10.1088/1361-6382/aabcb6} {\bibfield  {journal} {\bibinfo  {journal} {Class.
  Quant. Grav.}\ }\textbf {\bibinfo {volume} {35}},\ \bibinfo {pages} {115003}
  (\bibinfo {year} {2018})},\ \Eprint {http://arxiv.org/abs/1711.04538}
  {arXiv:1711.04538 [gr-qc]} \BibitemShut {NoStop}%
\bibitem [{\citenamefont {Novikov}\ and\ \citenamefont {Thorne}(1973)}]{ref70}%
  \BibitemOpen
  \bibfield  {author} {\bibinfo {author} {\bibfnamefont {I.~D.}\ \bibnamefont
  {Novikov}}\ and\ \bibinfo {author} {\bibfnamefont {K.~S.}\ \bibnamefont
  {Thorne}},\ }in\ \href@noop {} {\emph {\bibinfo {booktitle} {{Les Houches
  Summer School of Theoretical Physics}: {Black Holes}}}}\ (\bibinfo {year}
  {1973})\ pp.\ \bibinfo {pages} {343--550}\BibitemShut {NoStop}%
\bibitem [{\citenamefont {Page}\ and\ \citenamefont {Thorne}(1974)}]{ref79}%
  \BibitemOpen
  \bibfield  {author} {\bibinfo {author} {\bibfnamefont {D.~N.}\ \bibnamefont
  {Page}}\ and\ \bibinfo {author} {\bibfnamefont {K.~S.}\ \bibnamefont
  {Thorne}},\ }\href {\doibase 10.1086/152990} {\bibfield  {journal} {\bibinfo
  {journal} {Astrophys. J.}\ }\textbf {\bibinfo {volume} {191}},\ \bibinfo
  {pages} {499} (\bibinfo {year} {1974})}\BibitemShut {NoStop}%
\bibitem [{\citenamefont {{Bambi}}\ \emph {et~al.}(2021)\citenamefont
  {{Bambi}}, \citenamefont {{Brenneman}}, \citenamefont {{Dauser}},
  \citenamefont {{Garc{\'\i}a}}, \citenamefont {{Grinberg}}, \citenamefont
  {{Ingram}}, \citenamefont {{Jiang}}, \citenamefont {{Liu}}, \citenamefont
  {{Lohfink}}, \citenamefont {{Marinucci}}, \citenamefont {{Mastroserio}},
  \citenamefont {{Middei}}, \citenamefont {{Nampalliwar}}, \citenamefont
  {{Nied{\'z}wiecki}}, \citenamefont {{Steiner}}, \citenamefont {{Tripathi}},\
  and\ \citenamefont {{Zdziarski}}}]{2021SSRv..217...65B}%
  \BibitemOpen
  \bibfield  {author} {\bibinfo {author} {\bibfnamefont {C.}~\bibnamefont
  {{Bambi}}}, \bibinfo {author} {\bibfnamefont {L.~W.}\ \bibnamefont
  {{Brenneman}}}, \bibinfo {author} {\bibfnamefont {T.}~\bibnamefont
  {{Dauser}}}, \bibinfo {author} {\bibfnamefont {J.~A.}\ \bibnamefont
  {{Garc{\'\i}a}}}, \bibinfo {author} {\bibfnamefont {V.}~\bibnamefont
  {{Grinberg}}}, \bibinfo {author} {\bibfnamefont {A.}~\bibnamefont
  {{Ingram}}}, \bibinfo {author} {\bibfnamefont {J.}~\bibnamefont {{Jiang}}},
  \bibinfo {author} {\bibfnamefont {H.}~\bibnamefont {{Liu}}}, \bibinfo
  {author} {\bibfnamefont {A.~M.}\ \bibnamefont {{Lohfink}}}, \bibinfo {author}
  {\bibfnamefont {A.}~\bibnamefont {{Marinucci}}}, \bibinfo {author}
  {\bibfnamefont {G.}~\bibnamefont {{Mastroserio}}}, \bibinfo {author}
  {\bibfnamefont {R.}~\bibnamefont {{Middei}}}, \bibinfo {author}
  {\bibfnamefont {S.}~\bibnamefont {{Nampalliwar}}}, \bibinfo {author}
  {\bibfnamefont {A.}~\bibnamefont {{Nied{\'z}wiecki}}}, \bibinfo {author}
  {\bibfnamefont {J.~F.}\ \bibnamefont {{Steiner}}}, \bibinfo {author}
  {\bibfnamefont {A.}~\bibnamefont {{Tripathi}}}, \ and\ \bibinfo {author}
  {\bibfnamefont {A.~A.}\ \bibnamefont {{Zdziarski}}},\ }\href {\doibase
  10.1007/s11214-021-00841-8} {\bibfield  {journal} {\bibinfo  {journal} {Space
  Sci. Rev.}\ }\textbf {\bibinfo {volume} {217}},\ \bibinfo {eid} {65}
  (\bibinfo {year} {2021})},\ \Eprint {http://arxiv.org/abs/2011.04792}
  {arXiv:2011.04792 [astro-ph.HE]} \BibitemShut {NoStop}%
\bibitem [{\citenamefont {Zhang}\ \emph {et~al.}(1997)\citenamefont {Zhang},
  \citenamefont {Cui},\ and\ \citenamefont {Chen}}]{Zhang_1997}%
  \BibitemOpen
  \bibfield  {author} {\bibinfo {author} {\bibfnamefont {S.~N.}\ \bibnamefont
  {Zhang}}, \bibinfo {author} {\bibfnamefont {W.}~\bibnamefont {Cui}}, \ and\
  \bibinfo {author} {\bibfnamefont {W.}~\bibnamefont {Chen}},\ }\href {\doibase
  10.1086/310705} {\bibfield  {journal} {\bibinfo  {journal} {The Astrophysical
  Journal}\ }\textbf {\bibinfo {volume} {482}},\ \bibinfo {pages} {L155}
  (\bibinfo {year} {1997})}\BibitemShut {NoStop}%
\bibitem [{\citenamefont {McClintock}\ \emph {et~al.}(2014)\citenamefont
  {McClintock}, \citenamefont {Narayan},\ and\ \citenamefont
  {Steiner}}]{McClintock:2013vwa}%
  \BibitemOpen
  \bibfield  {author} {\bibinfo {author} {\bibfnamefont {J.~E.}\ \bibnamefont
  {McClintock}}, \bibinfo {author} {\bibfnamefont {R.}~\bibnamefont {Narayan}},
  \ and\ \bibinfo {author} {\bibfnamefont {J.~F.}\ \bibnamefont {Steiner}},\
  }\href {\doibase 10.1007/s11214-013-0003-9} {\bibfield  {journal} {\bibinfo
  {journal} {Space Sci. Rev.}\ }\textbf {\bibinfo {volume} {183}},\ \bibinfo
  {pages} {295} (\bibinfo {year} {2014})},\ \Eprint
  {http://arxiv.org/abs/1303.1583} {arXiv:1303.1583 [astro-ph.HE]} \BibitemShut
  {NoStop}%
\bibitem [{\citenamefont {Piotrovich}\ \emph {et~al.}(2023)\citenamefont
  {Piotrovich}, \citenamefont {Shablovinskaya}, \citenamefont {Malygin},
  \citenamefont {Buliga},\ and\ \citenamefont
  {Natsvlishvili}}]{Piotrovich2023}%
  \BibitemOpen
  \bibfield  {author} {\bibinfo {author} {\bibfnamefont {M.~Y.}\ \bibnamefont
  {Piotrovich}}, \bibinfo {author} {\bibfnamefont {E.~S.}\ \bibnamefont
  {Shablovinskaya}}, \bibinfo {author} {\bibfnamefont {E.~A.}\ \bibnamefont
  {Malygin}}, \bibinfo {author} {\bibfnamefont {S.~D.}\ \bibnamefont {Buliga}},
  \ and\ \bibinfo {author} {\bibfnamefont {T.~M.}\ \bibnamefont
  {Natsvlishvili}},\ }\href {\doibase 10.1093/mnras/stad2934} {\bibfield
  {journal} {\bibinfo  {journal} {Mon. Not. Roy. Astron. Soc.}\ }\textbf
  {\bibinfo {volume} {526}},\ \bibinfo {pages} {2596} (\bibinfo {year}
  {2023})},\ \Eprint {http://arxiv.org/abs/2309.12944} {arXiv:2309.12944
  [astro-ph.HE]} \BibitemShut {NoStop}%
\bibitem [{\citenamefont {{Shakura}}\ and\ \citenamefont
  {{Sunyaev}}(1973)}]{Shakura1973}%
  \BibitemOpen
  \bibfield  {author} {\bibinfo {author} {\bibfnamefont {N.~I.}\ \bibnamefont
  {{Shakura}}}\ and\ \bibinfo {author} {\bibfnamefont {R.~A.}\ \bibnamefont
  {{Sunyaev}}},\ }\href@noop {} {\bibfield  {journal} {\bibinfo  {journal}
  {\aap}\ }\textbf {\bibinfo {volume} {24}},\ \bibinfo {pages} {337} (\bibinfo
  {year} {1973})}\BibitemShut {NoStop}%
\bibitem [{\citenamefont {{Kong}}\ \emph {et~al.}(2014)\citenamefont {{Kong}},
  \citenamefont {{Li}},\ and\ \citenamefont {{Bambi}}}]{Kong14}%
  \BibitemOpen
  \bibfield  {author} {\bibinfo {author} {\bibfnamefont {L.}~\bibnamefont
  {{Kong}}}, \bibinfo {author} {\bibfnamefont {Z.}~\bibnamefont {{Li}}}, \ and\
  \bibinfo {author} {\bibfnamefont {C.}~\bibnamefont {{Bambi}}},\ }\href
  {\doibase 10.1088/0004-637X/797/2/78} {\bibfield  {journal} {\bibinfo
  {journal} {Astrophys. J.}\ }\textbf {\bibinfo {volume} {797}},\ \bibinfo
  {eid} {78} (\bibinfo {year} {2014})},\ \Eprint
  {http://arxiv.org/abs/1405.1508} {arXiv:1405.1508 [gr-qc]} \BibitemShut
  {NoStop}%
\bibitem [{\citenamefont {Fender}\ and\ \citenamefont {Belloni}(2004)}]{ref68}%
  \BibitemOpen
  \bibfield  {author} {\bibinfo {author} {\bibfnamefont {R.}~\bibnamefont
  {Fender}}\ and\ \bibinfo {author} {\bibfnamefont {T.}~\bibnamefont
  {Belloni}},\ }\href {\doibase 10.1146/annurev.astro.42.053102.134031}
  {\bibfield  {journal} {\bibinfo  {journal} {Ann. Rev. Astron. Astrophys.}\
  }\textbf {\bibinfo {volume} {42}},\ \bibinfo {pages} {317} (\bibinfo {year}
  {2004})},\ \Eprint {http://arxiv.org/abs/astro-ph/0406483}
  {arXiv:astro-ph/0406483} \BibitemShut {NoStop}%
\bibitem [{\citenamefont {Markoff}\ \emph {et~al.}(2005)\citenamefont
  {Markoff}, \citenamefont {Nowak},\ and\ \citenamefont {Wilms}}]{ref83}%
  \BibitemOpen
  \bibfield  {author} {\bibinfo {author} {\bibfnamefont {S.}~\bibnamefont
  {Markoff}}, \bibinfo {author} {\bibfnamefont {M.~A.}\ \bibnamefont {Nowak}},
  \ and\ \bibinfo {author} {\bibfnamefont {J.}~\bibnamefont {Wilms}},\ }\href
  {\doibase 10.1086/497628} {\bibfield  {journal} {\bibinfo  {journal}
  {Astrophys. J.}\ }\textbf {\bibinfo {volume} {635}},\ \bibinfo {pages} {1203}
  (\bibinfo {year} {2005})},\ \Eprint {http://arxiv.org/abs/astro-ph/0509028}
  {arXiv:astro-ph/0509028} \BibitemShut {NoStop}%
\bibitem [{\citenamefont {Mirabel}\ and\ \citenamefont
  {Rodriguez}(1999)}]{ref77}%
  \BibitemOpen
  \bibfield  {author} {\bibinfo {author} {\bibfnamefont {I.~F.}\ \bibnamefont
  {Mirabel}}\ and\ \bibinfo {author} {\bibfnamefont {L.~F.}\ \bibnamefont
  {Rodriguez}},\ }\href {\doibase 10.1146/annurev.astro.37.1.409} {\bibfield
  {journal} {\bibinfo  {journal} {Ann. Rev. Astron. Astrophys.}\ }\textbf
  {\bibinfo {volume} {37}},\ \bibinfo {pages} {409} (\bibinfo {year} {1999})},\
  \Eprint {http://arxiv.org/abs/astro-ph/9902062} {arXiv:astro-ph/9902062}
  \BibitemShut {NoStop}%
\bibitem [{\citenamefont {{Punsly}}\ and\ \citenamefont
  {{Coroniti}}(1990)}]{ref85}%
  \BibitemOpen
  \bibfield  {author} {\bibinfo {author} {\bibfnamefont {B.}~\bibnamefont
  {{Punsly}}}\ and\ \bibinfo {author} {\bibfnamefont {F.~V.}\ \bibnamefont
  {{Coroniti}}},\ }\href {\doibase 10.1086/168717} {\bibfield  {journal}
  {\bibinfo  {journal} {\apj}\ }\textbf {\bibinfo {volume} {354}},\ \bibinfo
  {pages} {583} (\bibinfo {year} {1990})}\BibitemShut {NoStop}%
\bibitem [{\citenamefont {Koide}(2003)}]{ref87}%
  \BibitemOpen
  \bibfield  {author} {\bibinfo {author} {\bibfnamefont {S.}~\bibnamefont
  {Koide}},\ }\href {\doibase 10.1103/PhysRevD.67.104010} {\bibfield  {journal}
  {\bibinfo  {journal} {Phys. Rev. D}\ }\textbf {\bibinfo {volume} {67}},\
  \bibinfo {pages} {104010} (\bibinfo {year} {2003})}\BibitemShut {NoStop}%
\bibitem [{\citenamefont {Blandford}\ and\ \citenamefont
  {Znajek}(1977)}]{ref69}%
  \BibitemOpen
  \bibfield  {author} {\bibinfo {author} {\bibfnamefont {R.~D.}\ \bibnamefont
  {Blandford}}\ and\ \bibinfo {author} {\bibfnamefont {R.~L.}\ \bibnamefont
  {Znajek}},\ }\href {\doibase 10.1093/mnras/179.3.433} {\bibfield  {journal}
  {\bibinfo  {journal} {Mon. Not. Roy. Astron. Soc.}\ }\textbf {\bibinfo
  {volume} {179}},\ \bibinfo {pages} {433} (\bibinfo {year}
  {1977})}\BibitemShut {NoStop}%
\bibitem [{\citenamefont {Tchekhovskoy}\ \emph {et~al.}(2010)\citenamefont
  {Tchekhovskoy}, \citenamefont {Narayan},\ and\ \citenamefont
  {McKinney}}]{ref89}%
  \BibitemOpen
  \bibfield  {author} {\bibinfo {author} {\bibfnamefont {A.}~\bibnamefont
  {Tchekhovskoy}}, \bibinfo {author} {\bibfnamefont {R.}~\bibnamefont
  {Narayan}}, \ and\ \bibinfo {author} {\bibfnamefont {J.~C.}\ \bibnamefont
  {McKinney}},\ }\href {\doibase 10.1088/0004-637X/711/1/50} {\bibfield
  {journal} {\bibinfo  {journal} {Astrophys. J.}\ }\textbf {\bibinfo {volume}
  {711}},\ \bibinfo {pages} {50} (\bibinfo {year} {2010})},\ \Eprint
  {http://arxiv.org/abs/0911.2228} {arXiv:0911.2228 [astro-ph.HE]} \BibitemShut
  {NoStop}%
\bibitem [{\citenamefont {Pei}\ \emph {et~al.}(2016)\citenamefont {Pei},
  \citenamefont {Nampalliwar}, \citenamefont {Bambi},\ and\ \citenamefont
  {Middleton}}]{ref90}%
  \BibitemOpen
  \bibfield  {author} {\bibinfo {author} {\bibfnamefont {G.}~\bibnamefont
  {Pei}}, \bibinfo {author} {\bibfnamefont {S.}~\bibnamefont {Nampalliwar}},
  \bibinfo {author} {\bibfnamefont {C.}~\bibnamefont {Bambi}}, \ and\ \bibinfo
  {author} {\bibfnamefont {M.~J.}\ \bibnamefont {Middleton}},\ }\href {\doibase
  10.1140/epjc/s10052-016-4387-z} {\bibfield  {journal} {\bibinfo  {journal}
  {Eur. Phys. J. C}\ }\textbf {\bibinfo {volume} {76}},\ \bibinfo {pages} {534}
  (\bibinfo {year} {2016})},\ \Eprint {http://arxiv.org/abs/1606.04643}
  {arXiv:1606.04643 [gr-qc]} \BibitemShut {NoStop}%
\bibitem [{\citenamefont {Daly}(2019)}]{Daly2019}%
  \BibitemOpen
  \bibfield  {author} {\bibinfo {author} {\bibfnamefont {R.~A.}\ \bibnamefont
  {Daly}},\ }\href {\doibase 10.3847/1538-4357/ab35e6} {\bibfield  {journal}
  {\bibinfo  {journal} {Astrophys. J.}\ }\textbf {\bibinfo {volume} {886}},\
  \bibinfo {pages} {37} (\bibinfo {year} {2019})},\ \Eprint
  {http://arxiv.org/abs/1905.11319} {arXiv:1905.11319 [astro-ph.HE]}
  \BibitemShut {NoStop}%
\bibitem [{\citenamefont {Daly}(2020)}]{Daly2020}%
  \BibitemOpen
  \bibfield  {author} {\bibinfo {author} {\bibfnamefont {R.~A.}\ \bibnamefont
  {Daly}},\ }\href {\doibase 10.1093/mnras/staa3213} {\bibfield  {journal}
  {\bibinfo  {journal} {Mon. Not. Roy. Astron. Soc.}\ }\textbf {\bibinfo
  {volume} {500}},\ \bibinfo {pages} {215} (\bibinfo {year} {2020})},\ \Eprint
  {http://arxiv.org/abs/2010.06908} {arXiv:2010.06908 [astro-ph.GA]}
  \BibitemShut {NoStop}%
\bibitem [{\citenamefont {Gou}\ \emph {et~al.}(2010)\citenamefont {Gou},
  \citenamefont {McClintock}, \citenamefont {Steiner}, \citenamefont {Narayan},
  \citenamefont {Cantrell}, \citenamefont {Bailyn},\ and\ \citenamefont
  {Orosz}}]{ref98}%
  \BibitemOpen
  \bibfield  {author} {\bibinfo {author} {\bibfnamefont {L.}~\bibnamefont
  {Gou}}, \bibinfo {author} {\bibfnamefont {J.~E.}\ \bibnamefont {McClintock}},
  \bibinfo {author} {\bibfnamefont {J.~F.}\ \bibnamefont {Steiner}}, \bibinfo
  {author} {\bibfnamefont {R.}~\bibnamefont {Narayan}}, \bibinfo {author}
  {\bibfnamefont {A.~G.}\ \bibnamefont {Cantrell}}, \bibinfo {author}
  {\bibfnamefont {C.~D.}\ \bibnamefont {Bailyn}}, \ and\ \bibinfo {author}
  {\bibfnamefont {J.~A.}\ \bibnamefont {Orosz}},\ }\href {\doibase
  10.1088/2041-8205/718/2/L122} {\bibfield  {journal} {\bibinfo  {journal}
  {Astrophys. J. Lett.}\ }\textbf {\bibinfo {volume} {718}},\ \bibinfo {pages}
  {L122} (\bibinfo {year} {2010})},\ \Eprint {http://arxiv.org/abs/1002.2211}
  {arXiv:1002.2211 [astro-ph.HE]} \BibitemShut {NoStop}%
\bibitem [{\citenamefont {Steiner}\ \emph {et~al.}(2012)\citenamefont
  {Steiner}, \citenamefont {McClintock},\ and\ \citenamefont {Reid}}]{ref100}%
  \BibitemOpen
  \bibfield  {author} {\bibinfo {author} {\bibfnamefont {J.~F.}\ \bibnamefont
  {Steiner}}, \bibinfo {author} {\bibfnamefont {J.~E.}\ \bibnamefont
  {McClintock}}, \ and\ \bibinfo {author} {\bibfnamefont {M.~J.}\ \bibnamefont
  {Reid}},\ }\href {\doibase 10.1088/2041-8205/745/1/L7} {\bibfield  {journal}
  {\bibinfo  {journal} {Astrophys. J. Lett.}\ }\textbf {\bibinfo {volume}
  {745}},\ \bibinfo {pages} {L7} (\bibinfo {year} {2012})},\ \Eprint
  {http://arxiv.org/abs/1111.2388} {arXiv:1111.2388 [astro-ph.HE]} \BibitemShut
  {NoStop}%
\bibitem [{\citenamefont {Steiner}\ \emph {et~al.}(2011)\citenamefont
  {Steiner}, \citenamefont {Reis}, \citenamefont {McClintock}, \citenamefont
  {Narayan}, \citenamefont {Remillard}, \citenamefont {Orosz}, \citenamefont
  {Gou}, \citenamefont {Fabian},\ and\ \citenamefont
  {Torres}}]{steiner2011spin}%
  \BibitemOpen
  \bibfield  {author} {\bibinfo {author} {\bibfnamefont {J.~F.}\ \bibnamefont
  {Steiner}}, \bibinfo {author} {\bibfnamefont {R.~C.}\ \bibnamefont {Reis}},
  \bibinfo {author} {\bibfnamefont {J.~E.}\ \bibnamefont {McClintock}},
  \bibinfo {author} {\bibfnamefont {R.}~\bibnamefont {Narayan}}, \bibinfo
  {author} {\bibfnamefont {R.~A.}\ \bibnamefont {Remillard}}, \bibinfo {author}
  {\bibfnamefont {J.~A.}\ \bibnamefont {Orosz}}, \bibinfo {author}
  {\bibfnamefont {L.}~\bibnamefont {Gou}}, \bibinfo {author} {\bibfnamefont
  {A.~C.}\ \bibnamefont {Fabian}}, \ and\ \bibinfo {author} {\bibfnamefont
  {M.~A.}\ \bibnamefont {Torres}},\ }\href@noop {} {\bibfield  {journal}
  {\bibinfo  {journal} {Monthly Notices of the Royal Astronomical Society}\
  }\textbf {\bibinfo {volume} {416}},\ \bibinfo {pages} {941} (\bibinfo {year}
  {2011})}\BibitemShut {NoStop}%
\bibitem [{\citenamefont {Chen}\ \emph {et~al.}(2016)\citenamefont {Chen},
  \citenamefont {Gou}, \citenamefont {McClintock}, \citenamefont {Steiner},
  \citenamefont {Wu}, \citenamefont {Xu}, \citenamefont {Orosz},\ and\
  \citenamefont {Xiang}}]{ref109}%
  \BibitemOpen
  \bibfield  {author} {\bibinfo {author} {\bibfnamefont {Z.}~\bibnamefont
  {Chen}}, \bibinfo {author} {\bibfnamefont {L.}~\bibnamefont {Gou}}, \bibinfo
  {author} {\bibfnamefont {J.~E.}\ \bibnamefont {McClintock}}, \bibinfo
  {author} {\bibfnamefont {J.~F.}\ \bibnamefont {Steiner}}, \bibinfo {author}
  {\bibfnamefont {J.}~\bibnamefont {Wu}}, \bibinfo {author} {\bibfnamefont
  {W.}~\bibnamefont {Xu}}, \bibinfo {author} {\bibfnamefont {J.}~\bibnamefont
  {Orosz}}, \ and\ \bibinfo {author} {\bibfnamefont {Y.}~\bibnamefont
  {Xiang}},\ }\href {\doibase 10.3847/0004-637X/825/1/45} {\bibfield  {journal}
  {\bibinfo  {journal} {Astrophys. J.}\ }\textbf {\bibinfo {volume} {825}},\
  \bibinfo {pages} {45} (\bibinfo {year} {2016})},\ \Eprint
  {http://arxiv.org/abs/1601.00615} {arXiv:1601.00615 [astro-ph.HE]}
  \BibitemShut {NoStop}%
\bibitem [{\citenamefont {{Shafee}}\ \emph {et~al.}(2006)\citenamefont
  {{Shafee}}, \citenamefont {{McClintock}}, \citenamefont {{Narayan}},
  \citenamefont {{Davis}}, \citenamefont {{Li}},\ and\ \citenamefont
  {{Remillard}}}]{Shafee06}%
  \BibitemOpen
  \bibfield  {author} {\bibinfo {author} {\bibfnamefont {R.}~\bibnamefont
  {{Shafee}}}, \bibinfo {author} {\bibfnamefont {J.~E.}\ \bibnamefont
  {{McClintock}}}, \bibinfo {author} {\bibfnamefont {R.}~\bibnamefont
  {{Narayan}}}, \bibinfo {author} {\bibfnamefont {S.~W.}\ \bibnamefont
  {{Davis}}}, \bibinfo {author} {\bibfnamefont {L.-X.}\ \bibnamefont {{Li}}}, \
  and\ \bibinfo {author} {\bibfnamefont {R.~A.}\ \bibnamefont {{Remillard}}},\
  }\href {\doibase 10.1086/498938} {\bibfield  {journal} {\bibinfo  {journal}
  {Astrophys. J. Lett}\ }\textbf {\bibinfo {volume} {636}},\ \bibinfo {pages}
  {L113} (\bibinfo {year} {2006})},\ \Eprint
  {http://arxiv.org/abs/astro-ph/0508302} {astro-ph/0508302} \BibitemShut
  {NoStop}%
\bibitem [{\citenamefont {McClintock}\ \emph {et~al.}(2006)\citenamefont
  {McClintock}, \citenamefont {Shafee}, \citenamefont {Narayan}, \citenamefont
  {Remillard}, \citenamefont {Davis},\ and\ \citenamefont {Li}}]{ref118}%
  \BibitemOpen
  \bibfield  {author} {\bibinfo {author} {\bibfnamefont {J.~E.}\ \bibnamefont
  {McClintock}}, \bibinfo {author} {\bibfnamefont {R.}~\bibnamefont {Shafee}},
  \bibinfo {author} {\bibfnamefont {R.}~\bibnamefont {Narayan}}, \bibinfo
  {author} {\bibfnamefont {R.~A.}\ \bibnamefont {Remillard}}, \bibinfo {author}
  {\bibfnamefont {S.~W.}\ \bibnamefont {Davis}}, \ and\ \bibinfo {author}
  {\bibfnamefont {L.-X.}\ \bibnamefont {Li}},\ }\href {\doibase 10.1086/508457}
  {\bibfield  {journal} {\bibinfo  {journal} {Astrophys. J.}\ }\textbf
  {\bibinfo {volume} {652}},\ \bibinfo {pages} {518} (\bibinfo {year}
  {2006})},\ \Eprint {http://arxiv.org/abs/astro-ph/0606076}
  {arXiv:astro-ph/0606076} \BibitemShut {NoStop}%
\bibitem [{\citenamefont {{Middleton}}\ \emph {et~al.}(2014)\citenamefont
  {{Middleton}}, \citenamefont {{Miller-Jones}},\ and\ \citenamefont
  {{Fender}}}]{ref94}%
  \BibitemOpen
  \bibfield  {author} {\bibinfo {author} {\bibfnamefont {M.~J.}\ \bibnamefont
  {{Middleton}}}, \bibinfo {author} {\bibfnamefont {J.~C.~A.}\ \bibnamefont
  {{Miller-Jones}}}, \ and\ \bibinfo {author} {\bibfnamefont {R.~P.}\
  \bibnamefont {{Fender}}},\ }\href {\doibase 10.1093/mnras/stu056} {\bibfield
  {journal} {\bibinfo  {journal} {mnras}\ }\textbf {\bibinfo {volume} {439}},\
  \bibinfo {pages} {1740} (\bibinfo {year} {2014})},\ \Eprint
  {http://arxiv.org/abs/1401.1829} {arXiv:1401.1829 [astro-ph.HE]} \BibitemShut
  {NoStop}%
\bibitem [{\citenamefont {Tchekhovskoy}\ \emph {et~al.}(2011)\citenamefont
  {Tchekhovskoy}, \citenamefont {Narayan},\ and\ \citenamefont
  {McKinney}}]{Tchekhovskoyetal.2011}%
  \BibitemOpen
  \bibfield  {author} {\bibinfo {author} {\bibfnamefont {A.}~\bibnamefont
  {Tchekhovskoy}}, \bibinfo {author} {\bibfnamefont {R.}~\bibnamefont
  {Narayan}}, \ and\ \bibinfo {author} {\bibfnamefont {J.~C.}\ \bibnamefont
  {McKinney}},\ }\href {\doibase 10.1111/j.1745-3933.2011.01147.x} {\bibfield
  {journal} {\bibinfo  {journal} {Monthly Notices of the Royal Astronomical
  Society: Letters}\ }\textbf {\bibinfo {volume} {418}},\ \bibinfo {pages}
  {L79} (\bibinfo {year} {2011})}\BibitemShut {NoStop}%
\bibitem [{\citenamefont {Russell}\ \emph {et~al.}(2013)\citenamefont
  {Russell}, \citenamefont {Gallo},\ and\ \citenamefont
  {Fender}}]{Russell2013ws}%
  \BibitemOpen
  \bibfield  {author} {\bibinfo {author} {\bibfnamefont {D.~M.}\ \bibnamefont
  {Russell}}, \bibinfo {author} {\bibfnamefont {E.}~\bibnamefont {Gallo}}, \
  and\ \bibinfo {author} {\bibfnamefont {R.~P.}\ \bibnamefont {Fender}},\
  }\href {\doibase 10.1093/mnras/stt176} {\bibfield  {journal} {\bibinfo
  {journal} {Mon. Not. Roy. Astron. Soc.}\ }\textbf {\bibinfo {volume} {431}},\
  \bibinfo {pages} {405} (\bibinfo {year} {2013})},\ \Eprint
  {http://arxiv.org/abs/1301.6771} {arXiv:1301.6771 [astro-ph.HE]} \BibitemShut
  {NoStop}%
\end{thebibliography}%

\end{document}